\documentclass[journal,twoside,final]{IEEEtran}
\usepackage{caption,subcaption} 
\usepackage{flushend}
\usepackage{orcidlink}
\usepackage{array, rotating}
\usepackage{makecell}
\usepackage{multicol,multirow, blindtext}
\usepackage{cite}
\usepackage{amsmath,amssymb,amsfonts}
\usepackage{graphicx}
\usepackage{ragged2e}
\usepackage{textcomp}
\usepackage{mathtools}
\usepackage[utf8]{inputenc}
\usepackage[linesnumbered,ruled,vlined]{algorithm2e}
\usepackage{algpseudocode}
\usepackage{textcomp}
\usepackage{booktabs}
\usepackage{mathrsfs}
\usepackage{cite}
\usepackage{setspace}
\usepackage{mathtools, cuted} 
\usepackage{url}
\usepackage[flushleft]{threeparttable}
\usepackage[font=small]{caption}

\usepackage[normalem]{ulem}  % \sout
\usepackage{color} 
\usepackage{ifthen}

\newboolean{trackchanges}
\setboolean{trackchanges}{true}    %<<<<<<<<<<<<  TRACK  <<<<<<<<<<<<<<<<
\ifthenelse{\boolean{trackchanges}}
{

	\newcommand{\del}[1]{\sout{\color{blue}#1}} 
}
{

	\newcommand{\del}[1]{}
}

\normalsize

\begin{document}
\graphicspath{{Images/}} 
\title{Semi-Contention-Free Access in IoT NOMA Networks: A Reinforcement Learning Framework}

\author{Abhishek Kumar$^{\orcidlink{0000-0003-1854-4842}}$,
José-Ramón Vidal$^{\orcidlink{0000-0002-7137-1349}}$, 
Jorge Martinez-Bauset$^{\orcidlink{0000-0003-3342-3037}}$,  
and Frank Y. Li$^{\orcidlink{0000-0003-4812-6211}}$ 
%$,~\IEEEmembership{Member,~IEEE}  
\vspace{-3mm}  
	
\thanks{Manuscript received February 18, 2025; revised June 6, 2025 and August 12, 2025, accepted September 12, 2025. The research leading to these results has received funding from the European Economic Area (EEA) Norway (NO) Grants 2014-2021, 
under Project contract no. 42/2021, RO-NO-2019-0499 --``A Massive MIMO Enabled IoT Platform with Networking Slicing for Beyond 5G IoV/V2X and Maritime Services (SOLID-B5G). 
The work of José-Ramón Vidal and Jorge Martinez-Bauset was also supported by Grant PID2021-123168NB-I00, funded by MCIN/AEI, 
Spain/10.13039/501100011033 and the European Union A way of making Europe/ERDF.
(Corresponding author: Frank Y. Li.)}
\thanks{Abhishek Kumar and Frank~Y.~Li are with the Department of Information and Communication Technology, University of Agder (UiA), N-4898 Grimstad, Norway (email: \{abhishek.kumar; frank.li\}@uia.no).} 
\thanks{José-Ramón Vidal and Jorge Martinez-Bauset are with the Departamento de Comunicaciones, Universitat Polit\`ecnica de Val\`encia (UPV), 46022 Val\`encia, Spain (email: \{jrvidal; jmartinez\}@upv.es).}
} 
\markboth{IEEE Transactions on Communications, Vol. XX, No. YY, 2025}
{Kumar \lowercase{\emph{et al}}.: Semi-Contention-Free Access in {IoT NOMA} Networks}
\maketitle
\begin{abstract} 
The unprecedented surge of massive Internet of things (mIoT) traffic in beyond fifth generation (B5G) communication systems calls for transformative approaches
for multiple access and data transmission. While classical model-based tools have been proven to be  powerful and precise, %for medium access and performance analysis,
an imminent trend for resource management in B5G networks is promoting solutions %heading swiftly 
towards data-driven design. %strategies. %considering network size and traffic uncertainty of IoT traffic. 
Considering an IoT network with devices spread in clusters covered by a base station, we present in this paper a novel model-free multiple access and data transmission framework empowered by reinforcement learning, designed for power-domain non-orthogonal multiple access networks to facilitate %data transmission for 
uplink traffic of small data packets.
%non orthogonal random access of uplink IoT traffic
%dedicated to clustered non-orthogonal multiple access-facilitated uplink small data traffic. 
The framework supports two access modes referred to as contention-based and semi-contention-free, 
with its core component being a policy gradient algorithm executed at the base station.
%for optimal access control and slot allocation. 
The base station performs access control and optimal radio resource allocation by periodically broadcasting two control parameters
%an access probability 
to each cluster of devices 
%and efficiently allocating radio resources 
that considerably reduce data detection failures with a minimum computation requirement on devices.
Numerical results, in terms of system and cluster throughput, throughput fairness, access delay, and energy consumption, 
demonstrate the efficiency and scalability of the framework as network size and traffic load vary.  
\end{abstract}
%\vspace{-3mm} 

\IEEEpeerreviewmaketitle
%===============================================================
\section{Introduction}
%===============================================================
\IEEEPARstart{A}{long} with the intensive deployment of fifth generation (5G) mobile networks worldwide focusing on enhanced mobile broadband (eMBB) services, ubiquitous connectivity in the form of 
%massive machine-type communications (mMTC) or
massive Internet of things (mIoT) heralds a transformative evolution towards sixth generation (6G)\cite{IoT_survey22}. 
According to ITU \cite{ITU-R_M2160}, the research target of connection density could reach $10^6\sim 10^8$ devices per square kilometer in 2030. 
 
Although a huge amount of IoT devices may be deployed across a cell, most of them are battery-powered devices and require typically infrequent data reporting or data collection~\cite{TGCN21}. 
%to transfer short length data packets sporadically.
%most of them do not perform data transmission frequently due to the constraint of energy consumption for battery-powered devices and event-driven feature of IoT data transmission. 
Indeed, IoT traffic exhibits a characteristic of \emph{small data} in terms of short packet size and %low traffic intensity as well as typically 
sporadic data transmissions. 
In the presence of overlapping transmissions based on the same radio resource, network performance may be degraded due to the reception of multiple signals that make the correct detection of individual signals impossible. %to be decoded correct
%on the same radio resources. transmission impairments that make signals impossible to be decoded correctly. 
Among various mechanisms to diminish data detection failures,
%caused by overlapping transmissions and to facilitate concurrent medium access, 
non-orthogonal multiple access (NOMA) and clustering are probably the two most eminent techniques. Through power-domain NOMA, concurrent transmissions from multiple users on %that contend for accessing 
the same radio resource can be disentangled and one or multiple transmissions may be successfully detected
%potentially deemed to be successful
through successive interference cancellation (SIC). On the other hand, clustering 
%on the can leverage spatial diversity through beamforming. 
which may be facilitated by multiple-input–multiple-output (MIMO)-based beamforming and beam steering can mitigate interference generated by users from other clusters~\cite{5G_book}. 
By jointly performing NOMA and clustering, the number of devices covered by the same beam could be greatly reduced and the benefit of SIC would be more significant. 

However, dealing with a substantial number of simultaneous
competing transmissions in clustered NOMA networks is not an easy task as both intra-cluster and inter-cluster interference\footnote{While intra-cluster interference indicates the interference generated by concurrent transmissions from other devices in the same cluster, inter-cluster interference is caused by the transmissions from any other cluster(s)\cite{IoTJ24}.} exists and no exact information on instantaneous access demands %traffic status 
for uplink IoT traffic is known to a base station (BS). 
Furthermore, how to assess the behavior of such networks analytically %theoretically 
and to design access schemes to optimize network performance \emph{in a real-time manner} call for further research endeavors, 
%remain largely as an un-chartered research topic, 
\emph{especially when network size grows}.  

%In network scenarios with lower device density or For small-scale networks, \emph{model-based} mathematical analysis using for instance information theory or Markov modeling %or Markov decision process
To assess the performance of a network, %small-scale networks, %a network operating with a fixed load, 
analytical models for example those that are built based on Markov chains have been proven to be a powerful tool, as they lead to closed-form expressions for various performance metrics%\cite{TCOM19}
~\cite{TCOM21}~\cite{IoTJ24}. However, such analytical %model-based
solutions may face scalability difficulties with respect to network size and complexity. On the other hand, the recent advances in artificial intelligence (AI) and machine learning (ML) enable numerous \emph{data-driven} approaches that envisage great potential to both providing real-time adaptation to dynamic network conditions and solving the scalability problem. 
In addition to the surge of interests in academia, the 3rd generation partnership project (3GPP) is also promoting ML applications in 6G radio access networks~\cite{TR38743}. Among a catalog %multiple types %a glossary %taxonomy 
of various ML approaches, reinforcement learning (RL)~\cite{RL_book} appears as a promising category to address the network scenario envisaged in this study.
%, as RL is \emph{model-free} without the need for external datasets for training and it improves network performance in a \emph{trial-and-error} manner~\cite{RL_book}.

\begin{table*}[ht]\small
\caption{Transposed comparison of learning-based clustering, access, and transmission schemes for uplink traffic in NOMA networks}
\centering
\renewcommand{\arraystretch}{1.2}
\setlength{\tabcolsep}{3pt}
%{\color{blue}
\begin{tabular}{|p{18mm}|p{24mm}|p{24mm}|p{24mm}|p{24mm}|p{24mm}|p{24mm}|}
\hline
\textbf{Category} & \textbf{Our schemes} & \textbf{\cite{rajasekaran2023neural}} & \textbf{\cite{TWC21}} & \textbf{\cite{ren2019em}} & \textbf{\cite{liu2021uplink}} & \textbf{\cite{IoTJ20}} \\
\hline
\textbf{Approach} & PG-based model-free access control and transmission for small data & ANN trained on heuristic clustering & Hybrid SARSA and DRL framework using 3D state actions & EM-based probabilistic clustering with online update & Optimization-based power and user allocation & LSTM-based DRL for sub-channel and power level selection \\
\hline
\textbf{Clustering Strategy} & Location-based clustering and probabilistic user pairing & SIC capability-aware user partitioning & Joint user, BS, and sub-channel mapping via state learning & Location-aware Gaussian mixture via unsupervised learning & Clustered access for hybrid spectrum sharing & Device sub-channel clustering to reduce collisions \\
\hline
\textbf{Learning Algorithm} & Reinforcement learning (policy gradient) & Offline supervised ANN & SARSA (light) + DQN (dense) & Statistical model-based (EM + online EM) & Non-learning (convex optimization) & DQN with LSTM  \\ \hline %(POMDP) \\ 
\textbf{System Type} & Uplink mIoT-NOMA & mmWave-NOMA & NOMA-IoT uplink & Downlink mmWave NOMA & Cognitive IoT uplink NOMA & Grant-free NOMA uplink \\
\hline
\textbf{Key Contributions} & RL-driven unified access control for CB and SCF modes; low device complexity & Heuristics for label generation and inference in real-time & Traffic-aware switching between RL types & Fixed \& dynamic user clustering; reduced re-clustering complexity & Support of both PU-first and PU-last decoding orders & Learning-based contention patterns for access success \\
\hline
\textbf{Performance Highlights} & Scalable system/ cell throughput; low delay; and efficient energy consumption & Heuristic performance with faster decision latency & Effective handling of dynamic traffic while maintaining fairness & Full EM performance while reducing runtime for dynamic scenarios & Balanced primary user protection and system throughput under hybrid access & Effective access policy and reduced contention over time \\
\hline
\end{tabular} \vspace{-3mm}
\label{tab:comparison}
\end{table*}

In the paper, we study a communication network where IoT devices
perform \emph{uplink random access and data transmission} in a hybrid semi-contention-free (SCF) or/and contention-based (CB) manner without requiring pre-allocated radio resources. From this perspective, %In this sense, 
the proposed access modes can be considered as a class of \emph{grant-free} (GF) random access. 
We regard the SCF access mode as an enhanced access mode
%combination of deterministic and random access 
that substantially improves the performance of conventional CB access by reducing both intra- and inter-cluster interference. 
 
As described latter, an RL-agent located at the BS performs access control and determines radio resource allocation policies, 
based solely on its observation of the number of successful \emph{small data packet} (SDP) detections in the uplink traffic. 
As an additional advantage, the proposed SCF access mode requires minimal computation capability on devices. 
The access schemes proposed in this study resemble the principle of \emph{random access small data transmission} described in \cite{sdt}, 
however, with much more advanced features facilitated by RL-driven access control and radio resource allocation.

%===============================================================
\subsection{Related Work} \label{subsec:related_work}
%===============================================================
%
%----------------------------------------------------------------------- 
\subsubsection{NOMA, clustering, GF/semi-GF (SGF)}% 
%----------------------------------------------------------------------- 
%In the literature, grant-free schemes are regarded as being more pragmatic for IoT traffic and numerous access schemes have been proposed
Among various access and data transmission schemes, many studies considered NOMA and clustering as promising bases for their scheme design.  
While NOMA can facilitate concurrent transmissions, clustering can further exploit the benefits of NOMA through device pairing \cite{GFIoT_survey20}. 
%(see \cite{GFIoT_survey20} and the references therein). %In~\cite{CL2022}, a NOMA-based access scheme in the presence of emergency and regular devices was proposed. 
In~\cite{CL2022}, a priority access NOMA-based slotted
ALOHA scheme was proposed, %for massive IoT access. The scheme deploys 
supporting multiple power levels and %associates devices to multiple 
priority levels. % each characterized by a specific distribution of %the use of the power levels. %to share the uplink to the BS. 
However, their proposed traffic estimator relays on
the existence of certain degree of symmetry between uplink
and downlink traffic %, and other assumptions 
which might not be realistic for IoT traffic.

Considering both intra- and inter-cluster interference in NOMA transmission, joint user clustering and power control for uplink traffic were studied in~\cite{TWC2022}. 
Recently, a contention-based coded random access scheme for heterogeneous traffic that allows both time and frequency domain resource sharing was proposed in~\cite{JSASP24}. 
Although these two studies investigated important
\emph{physical layer aspects}, they did not consider a dynamic traffic
scenario where random traffic is generated by devices, nor did they evaluate performance parameters such as throughput, delay, or fairness. 
Furthermore, we clarify that investigating such physical layer techniques is beyond the scope of this paper. 

%Given the fact that both GB and GF may apply to NOMA IoT data transmissions and each mode has its own advantages, another mode of access schemes, known as SGF, emerged~\cite{TCOM19}. 
%SGF schemes allow one or more devices enjoy GB access privilege while instructing other devices to compete for channel access according to the GF principle, leading to significant performance improvement in terms of outage probability \cite{TCOM21SGF}. 
In~\cite{TCOM19}, an access mode referred to as SGF was proposed, allowing GF devices share the dedicated
radio resources allocated to grant-based (GB) devices by exploiting NOMA and SIC. 
In an additional study, the authors proposed an enhanced SGF scheme to guarantee certain degree of quality of service (QoS) to GB devices~\cite{TCOM21SGF}.
Another recent paper proposed a novel SGF access scheme for short packet transmission by improving device-to-slot allocations based on the partial information observed by a BS~\cite{OJCOMSOC23}. 
%However, these SGF schemes were developed based on an assumption that \emph{the channel state information (CSI) %of the GB device(s) is available at the BS prior to data transmission}. %and such an assumption may not hold for IoT traffic. 
%Furthermore, the acquisition and processing of the CSI for multiple contending devices might be far beyond the hardware complexity and processing power of the devices considered in this study.
These proposals were developed based on an assumption
that either the BS or the GF users can acquire perfect
knowledge of the channel state information (CSI). However,
such an assumption seems to be unrealistic %difficult to achieve 
in real-life scenarios, particularly when a massive number of IoT devices contend for access and when considering the randomness of IoT data traffic~\cite{JSASP24}.
%----------------------------------------------------------------------- 
\subsubsection{Data-driven learning-based access schemes}
%----------------------------------------------------------------------- 
To design access and transmission schemes and analyze the performance of NOMA-enabled concurrent data transmissions require novel techniques. %For small-scale networks, classical \emph{model-based} tools including probability theory, optimization, and Markov decision process have been demonstrated to be powerful and precise as they provide explicit models with closed-form mathematical expressions for various performance metrics \cite{TCOM21}\cite{IoTJ24}\cite{TCOM21SGF}.  
%When access control is introduced, manual tuning or exhaustive searching has been a typical method for identifying optimal access probabilities. However, such approaches do not scale and are not applicable to large-scale networks.   In the meantime, 
As a trend beyond various conventional \emph{model-based} methods, there is a surge of endeavors in recent years to explore 
\emph{data-driven} learning-based approaches including RL-enabled access to %cluster-based 
uplink data transmissions as well as their potential benefits. %\cite{IoTJ20}. 

In \cite{rajasekaran2023neural}, two artificial neural network (ANN) algorithms that assign users to clusters in millimeter wave-NOMA (mmWave-NOMA) networks in a real-time manner were proposed. 
In~\cite{TWC21}, an RL-based resource allocation scheme was developed for uplink data transmissions in clustered NOMA IoT networks using deep reinforcement learning (DRL) and state–action–reward–state–action (SARSA)-learning algorithms based on three-dimensional (3D) state and action associations. In \cite{ren2019em}, an expectation maximization (EM)-based online clustering algorithm which is able to update user clusters through unsupervised learning based on location-awareness was developed. Furthermore, \cite{liu2021uplink} proposed a NOMA-based hybrid spectrum access scheme for uplink cognitive IoT traffic, supporting both decoding-primary-user(PU)-last and decoding-PU-first optimization.

In~\cite{access22}, a Q-learning (QL)-based random access method for NOMA IoT networks was proposed. 
A deep RL algorithm for throughput enhancement in GF NOMA systems was proposed in~\cite{IoTJ20} and two distributed QL algorithms for GF uplink transmissions in the presence of bursty IoT traffic were proposed in~\cite{IJSAC21}.
Moreover, a deep RL-based learning access scheme for signature-based GF transmissions was developed in~\cite{TCOM23}, 
targeting at maximizing long-term successful transmissions in NOMA-enabled beyond 5G (B5G) networks.   
%In \cite{TWC23}, a federated RL-based approach was proposed for distributed resource allocation, addressing privacy and scalability in massive IoT networks. 
Similarly,~\cite{Access23} introduced a double deep QL algorithm for efficient resource allocation in NOMA networks that handles dynamic traffic patterns. 
Another study in~\cite{IoTJ22} explored multi-agent RL techniques for enhancing uplink GF NOMA performance, resulting in improved throughput and fairness. Moreover, \cite{CL22} presented a novel RL-based power control mechanism combined with GF access to minimize energy consumption and maximize system reliability in clustered IoT networks. 
In a preliminary phase of this study, we proposed an RL-based random access scheme for IoT traffic where actions for access control were performed through QL~\cite{PIMRC24}. 
The aforementioned efforts to a certain extent unveil the promising role that RL may play in overcoming the complexities of medium access and data transmissions in B5G networks. In Table \ref{tab:comparison}, we provide a transposed comparison of our schemes with five representative schemes related to learning-based clustering, access, and transmissions in NOMA-enabled networks for IoT uplink traffic.     
%further reinforce the significance of RL-driven solutions in overcoming the complexities of massive IoT deployments in next-generation communication systems.

However, a number of research questions need to be answered before an RL-based approach can be applied to real-life B5G/6G networks. 
Among those, the following four questions triggered our motivation to perform the study reported in this paper. 
%1) For data transmission scheme design, can we combine contention-free access with contention-based access by considering the quality of service (QoS) requirement of users? 
1) What are the performance benefits that can be obtained by allowing 
multiple devices that perform concurrent SDP transmissions based on SCF and CB access share uplink radio resources?   
%uplink radio resources to be shared simultaneously by devices deploying SCF and CB access modes for SDP transmissions? 
%What are the performance benefits for devices deploying semicontention-free access when they share uplink resources with contention-based access devices.  
%2) With partial knowledge on network status based on its observation, how can a BS perform optimal resource allocation in a real-time manner without performing dedicated device-BS handshake on CSI?  
2) How can a BS determine high performance access control policies given that 
it can only observe the number of successful SDP transmissions and 
it does not know the number of devices simultaneously attempting to access common uplink radio resources?    
3) How close is the system performance achieved by the access control policies computed by an RL algorithm fed with partial system information from the ideal %optimal 
system performance? %In particular, when the BS does not know the number of active devices willing to access an uplink frame, and it only observes the MUD successes and failures of the uplink traffic. 
%3) How close can RL-based access control reach the theoretical performance limit where the BS retains exact knowledge on traffic status; %3) Based on the same network scenario and configuration, can one RL algorithm perform better than others? 
and 4) %Does the RL-based access scheme scale with network size?
How does the complexity and performance of an RL-based access scheme scale with network size and traffic load?

%==================================================================
%\vspace{-3mm}
\subsection{Contributions}
%==================================================================
When concurrent transmissions of two or more devices from one or multiple clusters occur on the same radio resource, the intricacies of SIC detection surge, as 
SIC alone may not resolve a collision when the obtained signal-to-interference-and-noise ratio (SINR) is not large enough \cite{IoTJ24}. 
To minimize SDP detection failures, a BS may impose access control to devices in order to %restrict their
limit the number of concurrent access to uplink radio resources. 
%However, \emph{the BS does not have exact knowledge on whether a device has a packet to transmit or not at a given time instant}. %Nor is CSI for IoT traffic always available at the BS. 
%Furthermore, devices may have distinct QoS requirements. 

Among various existing studies on NOMA-enabled IoT data transmissions, both \emph{saturated} and \emph{non-saturated} traffic conditions have been considered \cite{JSASP24}\cite{IoTJ20}. 
In the saturated case, the BS knows that all devices always have a packet to transmit~\cite{IoTJ20}. 
In the non-saturated case, where traffic state is unknown to the BS, packet arrival \emph{prediction} or \emph{estimation} 
serves as the basis for the design of an access control scheme~\cite{TCOM21}. 
However, such prediction schemes often rely on assumptions like Poisson arrivals or traffic models that may not hold in reality.  
     
In this paper, we introduce a random access and SDP transmission framework for uplink IoT traffic, 
referred to as RL for semi-contention-free (RL4SCF), and propose a \emph{data-driven model-free} RL-enabled access control and radio resource allocation mechanism  
that supports both the SCF and the CB random access modes. 
The operation of RL4SCF is \emph{observation-based}, without the need to assume any specific patterns of packet arrivals, %features of the traffic model,   
nor is it necessary for the BS to know the instantaneous traffic load within the cell. 
In addition, no specific signaling is required between the BS and the covered devices. 
Nor is it needed to perform any coordination among devices within a cluster or across different clusters.

%For access control, 
More specifically, we propose an RL-based policy gradient (PG)-driven access control and resource allocation algorithm, 
that allows the BS to compute and periodically broadcast an access probability to each cluster and a hash seed~\cite{RL_book}. 
In this way, access congestion is significantly reduced, as well as  inter- and intra-cluster interference.
In addition, the PG algorithm can be configured to improve throughput fairness among clusters located at different distances from the BS. 
%and evaluate its performance in comparison with the benchmark network performance which is obtained by assuming that the full knowledge on traffic state is available at the BS. 
As described later, the PG algorithm also computes \emph{hash seeds} to dynamically support the SCF access mode~\cite{TGCN21}. 
The hash seeds are computed to minimize the probability that the same uplink radio resource  
is simultaneously selected by multiple devices from the same cluster, significantly providing an additional reduction of both inter- and intra-cluster interference. 
Clearly, SDP detection failures have been substantially reduced through both access control and hash-based radio resource allocation, %time slot selection, 
leading to significant improvement on throughput, access delay, and energy consumption. 
%both the access control and the hash-based time slot selection features, substantially reduce interference and, therefore, the SDP detection failure rate, access delay and energy consumption are substantially improved. 

In brief, the novelty and main contributions of this paper are summarized as follows: 
 \begin{itemize}
\item A novel data-driven framework for clustered NOMA-enabled IoT that supports two uplink SDP access modes (SCF and CB) and 
%a BS observation-based 
an RL-enabled uplink access control and 
data transmission %resource management 
mechanism executed at the BS has been introduced. %, and supports two dedicated optimization goals, 
\item To support the SCF access mode, we devise a hash function-based %pseudo-random 
slot selection algorithm. 
Contending devices compute a hash function based on %that require 
two parameters, namely, the latest hash seed broadcasted by the BS 
and the device identity (ID).
The result identifies the time slot in a frame that the device shall use to transmit its SDP. 
No specific handshake between devices and the BS on channel condition or traffic state is needed.  
\item Relying on the \emph{partial system state} observed by the BS, a PG-driven algorithm has been proposed to dynamically adjust access probabilities and hash seed. 
The PG algorithm intends to maximize system throughput or cluster throughput fairness, 
relying only on its observation on the number of successfully detected SDP transmissions. 
%As \emph{online learning} progresses, the BS periodically broadcasts new updates computed by the PG algorithm. 
%
\item The framework has been implemented in a simulation platform. %using Python and Java respectively. 
Through extensive simulations, we validate the applicability and scalability of the developed framework with different network configurations 
and traffic load conditions, demonstrating that quasi-optimal system performance can be achieved. % by the PG algorithm. %despite partial knowledge.      
\item Our performance assessment is pursued from the traffic perspective. We aim at evaluating the impact that different 
network operation objectives %design goals 
have on four performance metrics: 
%based on four defined metrics, 
throughput, throughput fairness, access delay, and energy consumption. 
We also explore and shed light on the effectiveness and scalability of the proposed PG-enabled access control mechanism.%  framework. 
\end{itemize}

%{\color{blue}{\color{red}To be updated.}
In a nutshell, the uniqueness of this paper is represented by a combination of clustered NOMA-facilitated SDP transmission, 
the support of two %both SCF and CB 
access modes without bearing additional signaling overhead, %beyond the BS broadcast of the two parameters mentioned before, %contention-based and contention-free 
and RL-enabled access control through online learning. 
%The novelty of our proposal is the coordinated operation of access control and SCF mode, and how to run both simultaneously and efficiently, optimizing parameters in a real-time manner based on the BS’ partial observation on traffic state and device behavior.
The novelty of our proposal is represented by the coordinated operation of access control and radio resource allocation in a real-time manner supporting both CB and SCF access modes. The designed and implemented PG algorithm demonstrates that the RL-agent is able to efficiently and simultaneously learn, for each cluster, both an access control policy and a seed selection policy to minimize both intra- and inter-cluster interference.
To the best of our knowledge, the random access and data transmission framework developed in this paper, 
that is empowered by RL-enabled access control with multi-dimensional decision-making, 
is the first effort that tackles resource allocation for uplink IoT SDP transmission with a scalable RL-enabled % PG-driven 
solution that achieves quasi-optimal performance.       
 
The remainder of this paper is organized as follows. After presenting the envisaged network scenario and the physical layer transmission principle in Section~\ref{sec:system_model}, 
the RL-enabled framework for uplink IoT data transmission consisting of two access modes is introduced in Section~\ref{sec:framework}. 
Then the core component of the framework, a PG-driven access control mechanism is proposed in Section~\ref{sec:RL-agent} 
with its implementation overview outlined in Section \ref{sec:implementation}.
Section~\ref{sec:results} is dedicated %extensive simulations are performed 
to assess the performance of the developed framework %mechanism
with multiple network configurations and under various traffic load conditions. 
Furthermore, we discuss a few aspects that are related to the feasibility and operability of our framework in Section\ref{Sec:feasibility_discussions}.
Finally, the paper is concluded in Section~\ref{Sec:Conclusions}.

%{\color{red}A table of comparison will be added in this section.} 

%%%%%%%%%%%%%%%%%%%%%%%%%%%%%%
\begin{figure}
	\centering
	\includegraphics[scale = 0.43]{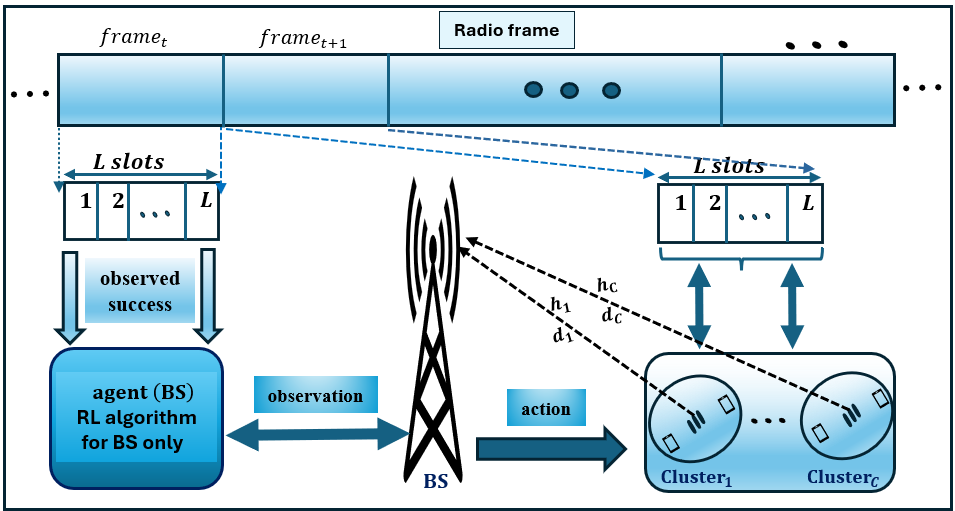} %\vspace{-1mm}
\caption{Overview of the RL4SCF framework: Network scenario, frame structure, data transmission, and RL-enabled access control.} \vspace{-3mm}
	\label{fig:framework}
\end{figure} 
%%%%%%%%%%%%%%%%%%%%%%%%%%%%%% 

%=========================================================================================== 
%\vspace{-1mm}
\section{Network Scenario and Transmission Principle}
\label{sec:system_model}
%=========================================================================================== 
In this section, we briefly present the network scenario and
the data transmission principle adopted in this study. 

%==================================================================
%\vspace{-1mm}
\subsection{Network Scenario}
%==================================================================
Consider a NOMA-enabled IoT network composed of a single cell with a beamforming-assisted BS, connecting numerous devices which are uniformly distributed across the cell. Devices in the studied network are battery-powered and equipped with a single antenna. An identical transmit power level applies to all SDP transmissions and no power control is introduced~\cite{sp}. For presentation simplicity, we show that a radio resource is perceived as structured in a framed slotted manner to facilitate random access and SDP transmission. 

In addition, the BS is assisted by a MIMO antenna connecting
numerous devices across the cell.
Through beamforming and beam steering, a group of devices with an (almost) identical angle toward the BS can be covered by the same beam, leading to a less number of devices covered by one beam \cite{5G_book}. To cover a whole cell across multiple directions,   
%will be greatly reduced since
distinct beams can be assigned. However, how to perform beamforming is beyond the scope of this paper. %to different groups of devices.

Furthermore, we consider that devices are static, i.e., they do not move, and assume that no device hardware or software failure occurs during the operation of the framework. Devices are distributed into multiple clusters, 
where each cluster confines a number of devices that are located in the vicinity of each other. 
As depicted in Fig.~\ref{fig:framework}, \textit{Cluster $i$} ($C_i$), $i = 1, \ldots, C$, is composed of $N_i$ devices and its center is located $d_i$ meters apart from the BS. %\footnote{{\color{blue}In our network scenario, the center of each cluster is located on a two-dimensional ground decided by $(x,y)$, and the location of the $j$-th device in cluster $C_i$ %, i.e., $(x_j^i,y_j^i)$, 
The locations of these devices may vary within the radius from the cluster center. %Then the height of the BS, $g$,   %which represents the third dimension $z$, is also needed to calculate the distance from a device to the BS. 
A cluster with a lower cluster index is located closer to the BS, e.g., $d_i < d_j$ if $i<j$. A summary of the main notations used in this paper can be found in the Appendix. 
%{\color{green} \sout{From the random access protocol (RAP) perspective, transmissions over time are organized into frames and each frame is composed of $L$ time slots of equal length. Along the text, when we refer to \emph{frame}, we are referring to an RL4SCF framework frame. This RL4SCF frame might not necessarily fit in a system frame, as defined in XXX.  }%\sout }%green

%----------------------------------------------------------------------- 
%\subsection{Transmission Principle}
\subsection{Frame Structure and Physical Layer Principles}
\label{subsec:transmission_principle}
%----------------------------------------------------------------------- 
%{\color{green} Along the text, when we refer to \emph{frame}, we are referring to an RL4SCF framework frame.}%green

\subsubsection{Frame structure}
To facilitate flexible and efficient radio resource allocation, multiple time slots are grouped into one frame. %A frame defined in this study consists of an integer number 
All time slots have the same duration and a device can select at most one of the time slots in a frame to transmit its SDP. Two phases, namely, SDP transmission and
acknowledgment (ACK) after a short interval upon a successful reception, occur inside one time slot. SIC is performed at the end of the data transmission phase and the ACK message is sent at the end of the time slot. %Moreover, we assume that the SDP size is smaller than \emph{a configured amount of UL data} so that no resource reservation is required before an SDP transmission~\cite{sdt}.   

%\del{Each device only transmits a single SDP per frame. However,} 
As concurrent transmissions from various devices located in the same or different clusters are allowed, one or multiple SDP transmissions from different devices may occur within one time slot. 
%green  

\subsubsection{Signal reception and channel gain}
The total received signal at the BS in the $k$-th time slot, $y_k$, 
%in a time slot, $y$ 
which is accumulated based on the individual signals from those devices that transmit concurrently within the same time slot, can be expressed as,
%\vspace{-2mm}

\begin{equation}
y_k = \sum\limits_{i=1}^C\sum\limits_{j= 1}^{N_i}\mathcal{I}\left(i,j,k\right)\textbf{H}_j^i x_j^i + \eta \,, 
\label{eq:received_signal}
%\vspace{-1mm}
\end{equation}
where $\mathcal{I}\left(i,j,k\right)$ is an indicator function that is $1$ when the $j$-th device in cluster $C_i$ transmits in the $k$-th time slot, and $0$ otherwise. 
In (\ref{eq:received_signal}), $\textbf{H}_j^i$, $x_j^i$, and $\eta$ represent 
the complex channel gain vector between device $D^i_j$ (the $j$-th device in the $i$-th cluster) and the BS, 
the transmitted signal by that device, and the additive noise %The additive noise present in the channel 
which follows a complex Gaussian distribution with zero mean and variance $\rho^2$, 
i.e., $\eta\sim{\mathcal{CN}} \left(0, \rho^2 \right)$, respectively. %{\color{blue} $\hat{N}_i$ represents that there are $\hat{N}_i$ devices from cluster $C_i$ that transmitted in this slot.}

%\begin{eqnarray}
%y = \sum\limits_{i=1}^C\sum\limits_{j= 1}^{\color{blue}{\hat{N}_i}}\textbf{H}_j^i x_j^i + \eta \,. 
%\label{eq:received_signal}
%\vspace{-1mm}
%\end{eqnarray}

%{\color{green} The total received signal at the BS in the $n$-th time slot $y_n$ can be expressed as,}%green  

%===============================================================================

\begin{table}[t] \footnotesize %\small
    \centering
    \caption{Successful data detection probabilities \cite{IoTJ24} }\vspace{-1mm}
    %{\color{blue} 
    \begin{tabular}{|p{8mm}|p{5mm}|p{5mm}|p{5mm}|p{5mm}||p{5mm}|p{5mm}|p{5mm}|p{5mm}|p{3mm}|}
    \hline
    \multirow{2}{*}{State}&\multicolumn{4}{c||}{Cluster $C_1$ }&\multicolumn{4}{c|}{Cluster $C_2$}\\
    \cline{2-9}
    ($n_1,n_2)$ &   $S_0$ & $S_1$ & $S_2$ & $S_3$  & $S_0$  & $S_1$ & $S_2$ & $S_3$\\
         \hline
    (0,0) & 1 & 0 & 0 & 0 & 1& 0 & 0 & 0 \\
    \cline{2-9} 
    \hline
    (0,1) & 1 & 0 & 0 & 0 & 0.567& 0.433 & 0 & 0 \\
    \cline{2-9} 
    \hline
    (0,2) & 1 & 0 & 0 & 0 & 0.477& 0.490 & 0.033 & 0 \\
    \cline{2-9} 
    \hline
    (0,3) & 1 & 0 & 0 & 0 & 0.521& 0.424 & 0.055 & 0 \\
    \cline{2-9} 
    \hline
    (1,0) & 0.163 & 0.837 & 0 & 0 & 1 & 0 & 0 & 0 \\
    \cline{2-9} 
    \hline
    (1,1) & 0.401 & 0.599 & 0 & 0 & 0.804 & 0.196 & 0 & 0 \\
    \cline{2-9} 
    \hline
    (1,2) & 0.564 & 0.436 & 0 & 0 & 0.766 & 0.227 & 0.007 & 0 \\
    \cline{2-9}  \hline
    (1,3) & 0.675 & 0.325 & 0 & 0 & 0.788 & 0.200 & 0.012 & 0 \\
    \cline{2-9}  \hline
    (2,0) & 0.606 & 0.274 & 0.120 & 0 & 1 & 0 & 0 & 0 \\
    \cline{2-9}  \hline
    (2,1) & 0.677 & 0.249 & 0.074 & 0 & 0.941 & 0.059 & 0 & 0 \\
    \cline{2-9}  \hline
    (2,2) & 0.734 & 0.220 & 0.046 & 0 & 0.929 & 0.070 & 0.001 & 0 \\
    \cline{2-9}  \hline
     (2,3) & 0.780 & 0.191 & 0.029 & 0 & 0.935 & 0.062 & 0.003 & 0 \\
    \cline{2-9}  \hline
    (3,0) & 0.741 & 0.167 & 0.079 & 0.013 & 1 & 0 & 0 & 0 \\
    \cline{2-9}  \hline
    (3,1) & 0.779 & 0.164 & 0.049 & 0.008 & 0.984 & 0.016 & 0 & 0 \\
    \cline{2-9}   \hline
    (3,2) & 0.810 & 0.155 & 0.031 & 0.004 & 0.980 & 0.020 & 0 & 0 \\
    \cline{2-9}  \hline
    (3,3) & 0.836 & 0.143 & 0.019 & 0.002 & 0.981 & 0.019 & 0 & 0 \\
    \cline{2-9}  \hline
     \end{tabular}
    \label{phy_layer}
    \vspace{-6mm}
\end{table}

The channel gain between devices and the BS, which is located $g$ meters above the ground,  
$\textbf{H}_j^i\sim{\mathcal{CN}} \left(\textbf{0}_M, \textbf{I}_M \right)$, follows a Rayleigh fading model with a zero mean complex Gaussian distribution, 
where $M \geq 1$ is the number of antennas mounted at the BS. 
Moreover, the signal transmitted by device $D_j^i$ is expressed as $x_j^i  = \sqrt{P}s_j^i $, 
where $P$ is the transmit power and 
$s_j^i$ is the transmitted signal %for device $D_j^i$ 
with unit variance. 
The path loss is determined by $128 + 37.6 \log_{10} d$, with $d$ being the distance (in kilometers) between a device and the BS in a Euclidean 3D space. 
%{\color{red}  \textbf{New expression for $d$} }%red   

%\del{Devices with a non-empty queue are referred to as \emph{active devices}. }
%{\color{red}  \textbf{defined later} }%red 
\subsubsection{Interference and data detection}
%Whether an SDP transmission is successful or not depends on the SINR level of the SDP signal with respect to the other interfering signals in the same time slot, and the noise level. 
The success or failure of an SDP transmission depends on the SINR level of the SDP signal relative to the other interfering signals in the same time slot and the noise level.
To potentially detect one SDP, the signals from all other concurrent transmissions within the same slot are treated as interference to the signal that has the strongest signal strength.
%the BS determines the SINR of each received signal, considering The signal with the strongest SINR will be decoded first
Applying the SIC principle for power-domain NOMA, one or more %of the 
SDP transmissions received by the BS in the same time slot may be detected successfully when their SINR is greater than a given threshold, $SINR = 10$ dB, which is a realistic threshold obtained from the real-life experiments performed in~\cite{access2021}. Upon a successful detection, the decoded packet is subtracted from the received signals and the detection procedure proceeds successively with one less signal until all signals are processed. 

Considering a cell with two clusters,
we have determined the probability distributions for successful detection when multiple packets from two clusters are transmitted in the same time slot. 
%In Table~\ref{phy_layer}, we illustrate these probabilities when one or up to six $C1$ and/or $C_2$ devices transmit concurrently, with the results reproduced from our early work~\cite{IoTJ24}.
%The obtained results are illustrated in Table~V of our early work~\cite{IoTJ24}. 
%These} distributions have the form $p_i(s,n_1,n_2)$, $i=1,2$\,, i.e., the probability that $s$ packets from cluster $C_i$ are successfully detected within a time slot that contains $(n_1, n_2)$ packets from clusters $C1$ and $C_2$, respectively. As an example, 
Table ~\ref{phy_layer} illustrates these distributions when $N_i =\{0,1,2, 3\}$, $i=1,2$\,, respectively. These results are reproduced from our earlier work~\cite{IoTJ24}, which applied the same distance and cluster radius configuration. 
Note that the table content for row $(n_1,n_2)$ and column $S_u$ is the probability that $u$ successful packet detections occur from the corresponding cluster, when the number of packets transmitted in the same time slot are $n_1$ and $n_2$ 
from $C_1$ and $C_2$, respectively. 
Take state $(n_1,n_2)=(2,2)$ as an example. When concurrent transmissions from four devices, two from each cluster, occur, 
there is a probability of 22\% of detecting one packet from $C_1$ and of 7\% of detecting one packet from $C_2$. Furthermore, 
the probabilities of two packets from the same cluster being detected successfully are 4.6\% and 0.1\% for $C1$ and $C_2$ devices respectively. Without NOMA, these probabilities would be a zero. Based on this observed benefit, we apply NOMA also in this study and reuse the probability distributions of successful detection shown in Table~\ref{phy_layer}. %also apply to this study.

%decoded from a time slot that contains $n_1$ packets from cluster $C_1$ and $n_2$ packets from cluster $C_2$. %{\color{blue}$p_i(n_1,n_2)$, i.e., the probability that $n_1$ and $n_2$ packets that are transmitted from cluster $C_1$ and $C_2$ respectively are successfully received within one time slot.} %{\color{red}The sentence in blue may need further update.}
%{\color{red}Table V in [6] has only two elements, as state $(n_1,n_2)$. So I think it is better not to use $p_i(s,n_1,n_2)$ here. For clusters, we denote them by $C_1$ and $C_2$ in the context of this paper.}

%{\color{red}This following points should be added in this section: 
%\begin{itemize}
%\item a short explanation on beamforming. No expression needed. 
%\item a short explanation on NOMA modeling. No expression needed. 
%\item a description on our frame structure. 
%\end{itemize} }

%----------------------------------------------------------------------- 
%\subsection{SDT}
%\label{subsec:transmission_principle}
%----------------------------------------------------------------------- 

%{\color{green} Active devices, i.e., those devices that have at least one packet ready for transmission, can send a maximum of one packet per frame. Conceptually, the RAP is composed of two schemes: i) the access control (permission) that runs at the BS; ii) the autonomous slot selection that runs at the devices.}%green  

\section{RL4SCF: An RL-enabled framework}
\label{sec:framework}
In this section, we present the developed framework as a whole and the constituent elements including access control, random access modes, hash function, and SDP transmission schemes. The RL algorithm that serves as the core for the operation of the framework will be presented later in Section~\ref{sec:RL-agent}. Four performance metrics are defined at the end of this section.

\subsection{Framework Overview}
An overview of the RL4SCF framework including the constituent elements and the operation of the framework is illustrated in Fig.~\ref{fig:framework}.
%both hardware and software {\color{blue}ingredients}. %compartments {\color{red}(maybe another word like compound, components, or elements)}. 
From a system composition perspective, the framework contains: 
1) a BS that performs not only conventional 5G functions but also the new mechanism and algorithm proposed in this study; and 
2) numerous IoT devices  confined in two (or multiple) clusters spread randomly across the cell coverage. 
From a network functionality perspective, the framework %features are: 
1) enhances the 5G functions by enabling RL at the BS for the purpose of achieving best possible %quasi-optimal
network performance (system throughput and cluster throughput fairness); and 
2) enables devices with CB or SCF access modes, supported through access control parameters periodically broadcasted by the BS.

\subsubsection{RL4SCF--BS operation} 
%------------------------------------------------------------------------
%\del{The behavior of the BS follows the procedure presented next.} 
%During the operation phase, the BS has no information on how many devices are active and among them how many will contend in a frame. 
The BS functions as an \emph{RL-agent}. Based on its observation on the number of successful SDP transmissions, the BS performs RL-based optimization, takes actions on the access probability for each cluster and the hash seed that supports the SCF access mode, and broadcasts the decisions periodically to all clusters.  
%in the form of periodic broadcasts of the access probability for each cluster and the hash seed to support the SCF access mode. 
%-----------------------------------------------------------------
\subsubsection{RL4SCF--device operation} 
%-----------------------------------------------------------------
Devices have a micro-controller unit (MCU) with simple computation capability. 
Each device operates independently from the operation of the other devices in the same cluster. 
%Devices are synchronized with the BS so that periodic operations can be performed. %slotted ALOHA can be performed. 
\emph{No RL capability} is required for the operation of devices. 
Nor is it needed to perform specific signaling prior to an SDP transmission by a device.

In brief, the RL4SCF framework includes two major components: 1) One access control and resource allocation \emph{mechanism} that is enabled by an RL \emph{algorithm}, executed at the BS; and 2) Two SDP transmission \emph{schemes} enabled by two random access \emph{modes}, executed by devices through the instructions by the BS. In  Table~\ref{tab:access_schemes}, we summarize how these components are integrated into the framework and the key features. The main elements 
of the framework are presented in the following subsections.

%-----------------------------------------------------------------------------
\begin{table}[t]
\renewcommand{\arraystretch}{1.3}
\small
\caption{Framework overview: Components, principles, \& features} 
\vspace{-1mm} 
\label{tab:access_schemes}
\centering \small
\resizebox{0.48\textwidth}{!}{%
%\begin{tabular}{|c|c|c|c|c|}
%{\color{blue}
\begin{tabular}{|p{21mm}|p{16mm}|p{40mm}|p{40mm}|}\hline
\bf{Component I:} & \multicolumn{3}{|c|} {\bf{RL-enabled access control and resource allocation}} \\ \hline   
Principle and features & \multicolumn{3}{|p{96mm}|} {Executed by the BS periodically and applies to access control and slot selection for devices in both clusters that have different access probabilities}  \\ \hline \hline 
\bf{Component II:} & \multicolumn{3}{|c|} {\bf{Random access and SDP transmission schemes}} \\ \hline   
Principle and features & Scheme A & CB mode for $C_1$ devices (slots selected
with equal probability) & CB mode for $C_2$ devices (slots selected
with equal probability)\\ \hline
Principle and features & Scheme B & SCF mode for $C_1$ devices (hash-based slot selection) & CB mode for $C_2$ devices (slots selected with equal probability) \\ \hline
\end{tabular}}% 
\vspace{-4mm}
\end{table}
%-----------------------------------------------------------------------------

%----------------------------------------------------------------------- 
 
\vspace{-2mm}
\subsection{Access Control}
%\label{subsec:transmission_principle}
%-----------------------------------------------------------------------
Let $W$ be the number of devices that transmit in a frame which contains $L$ time slots. Clearly, the number of successful transmissions within the frame could be higher than $L$, thanks to the enhanced data detection capability provided by NOMA. Nevertheless, for a given frame length $L$, there exists an optimum number of devices $W^* \ge L $ that can transmit in that frame. 
When $W > W^*$, the likelihood of occurring data detection failure %the number of collisions 
\emph{increases}, then the number of successful transmissions \emph{decreases}. %and energy is wasted. 
On the other hand, radio resources are underutilized when $W < W^*$.  
As a measure, imposing access control leads to optimal network performance.
%As a consequence, to deliver an optimal performance, the RAP must enforce a limit to $M$. 

As a vital component of the RL4SCF framework, we propose a probabilistic access control mechanism that determines $W^*$ \emph{per frame and per cluster}. 
The access control mechanism is facilitated by an RL algorithm that runs at the BS. As the RL-agent, the BS dynamically adapts its actions based on its observation on system state and periodically broadcasts an access probability $a_i$\footnote{Note that $a_i$ represents the fraction of active devices in $C_i$ that will transmit in the current frame.} to $C_i$.  
The aim of the RL algorithm is to make $W$ as close to $W^*$ as possible, while maximizing a given performance objective as defined later in Section \ref{sec:RL-agent}. 

%In summary, the operation of the random access protocol is as follows.

{\color{black}
Devices with a non-empty queue are referred to as \emph{active devices}. 
An active device in $C_i$ generates a random variate $\alpha$, uniformly distributed in $\left[0,1\right]$, 
and compares it with the latest received access probability $a_i$ from the BS.  
If $\alpha \leq a_i$\,,} it will transmit in the current frame and we refer to it as an \textit{active device} that \textit{transmits} (ADT).
Otherwise, we refer to it as an \textit{active device} forced to \textit{defer} (ADD) its transmission.

Within each frame, one active device may transmit \emph{a maximum of} one packet. 
Active devices follow the same procedure in every frame in a memory-less manner,
regardless of whether the next packet in the buffer is a newly arrived or a backlogged one. 
On the other hand, an ADD will attempt to
transmit in the next frame with the corresponding access probability received from the BS.
If a device transmits a packet but does not receive an ACK message, it will continue the same procedure until an ACK confirming the successful detection by the BS. 
%green 

%----------------------------------------------------------------------- 
\vspace{-2mm}
\subsection{Random Access Modes}
%\label{subsec:transmission_principle}
%----------------------------------------------------------------------- 
Once an active device has got access permission, it must select a single time slot in the frame to proceed with data transmission. 
%For this action we propose two algorithms: 
As explained below, two access modes %i) a simple random selection; ii) a hash function based selection, 
are defined in our framework and they constitute the basis for the SDP transmission schemes presented in the next subsection. 
 
\subsubsection{CB mode}
ADT devices perform simple random selection so that 
a single time slot is selected \emph{with equal probability} from the set of $L$ slots in a frame.  
%We refer to this random access mode that deploys a random selection as the CB access mode. 
This mode offers basic access. However, a \emph{slot selection collision} occurs if the same time slot in a frame is selected by multiple devices.  

Depending on whether the devices that select the same time slot are from different clusters or the same cluster, a slot selection collision could lead to more serious inter- or intra-cluster interference. From our previous studies on a similar network scenario~\cite{IoTJ24}\cite{PIMRC24}, with one representative result shown in Table \ref{phy_layer}, we conclude that intra-cluster interference appears as the most detrimental factor for NOMA-based data detection. The table also shows that avoiding slot selection collision from $C_1$ devices may improve $C_2$ SDP detections as well. 

\subsubsection{SCF mode}
In order to diminish data detection failures, %provide a service better than the basic one, we propose 
the SCF access mode is proposed. 
This mode is enabled by the execution of a hash function at the SCF devices.
The hash function aims to distribute the time slots selected by different devices using the SCF mode as widely as possible across the frame.
%The hash function achieves a goal that the time slots selected by different devices deploying the SCF mode are scattered across the frame as much as possible. 
As such, the number of slot selection collisions caused %by intra-cluster interference 
from the SCF devices is dramatically reduced.  
Clearly, the SCF mode provides an access service superior to the one provided by the CB mode, leading to improved throughput, shorter delay, and reduced energy consumption. 

The SCF mode is supported by an RL algorithm that runs at the BS and that periodically broadcasts a hash seed per cluster, e.g., seed $b_i$ to $C_i$. An ADT determines the transmission time slot by executing a hash function that requires a hash seed $b_i$ as input, as described next.

%Furthermore, the SCF mode is supported by an RL algorithm that is operated at the BS to select the best hash seed %an RL algorithm that runs at the BS 
%(to be presented in Section \ref{sec:RL-agent}) %and that periodically broadcasts a hash seed $b_i$ to $C_i$. 
%and the execution of the hash function at devices to determine the transmission time slot %requires a hash seed as input as described next. 
%(explained in the next subsection). %Subsection \ref{subsec:hash_function}).
%green 

%----------------------------------------------------------------------- 
\subsection{Hash-based Slot Selection}
\label{subsec:hash_function}
%----------------------------------------------------------------------- 
This element aims at avoiding or minimizing slot selection collision and reducing intra- and inter-cluster interference. 
Clearly, under heavy traffic conditions, this function cannot completely eliminate collisions, neither slot selection collisions nor data detection failures. 
Then, running hash-based function slot selection achieves the best performance when deployed together with an effective access control mechanism. 

For each cluster, the RL-agent at the BS keeps a set of candidate seeds and finds which are the best seeds in the set, given the current state of the network. To do this, the RL algorithm learns the expected performance obtained by each seed at each system state.
%when applied to the current set of active devices in the cluster. 
Note that the BS does not know the IDs of the ADTs. 
At a given frame, the best seed for each cluster will be the one that leads to the greatest number of successful transmissions. 
%From this, the RL algorithm chooses periodically a seed $b_i$ to be broadcasted to each cluster $C_i$.

The hash function $f_{hash}$ is executed by each ADT. As the input, it takes the latest hash seed broadcasted by the BS plus the unique ID of the ADT. As the output, it returns the slot number for SDP transmission in the corresponding frame:	
\begin{eqnarray}\label{eq:hashF}
slot= f_{hash}(b_i, \text{ID})\,,
\end{eqnarray} 
where $b_i$ is the seed broadcasted to cluster $C_i$. %\del{and $L$ is the number of time slots in one frame.}
As such, the slot chosen by each ADT is jointly determined by its ID and the hash seed.  
Then, each seed $b_i$ represents a mapping of the ADTs in cluster $C_i$ to the available slots.

For a given set of ADTs in $C_i$, the BS intends to find and broadcast a seed $b_i$ that results in more favorable mapping of devices to slots, i.e., with fewer slot selection collisions. 
Note that this set of ADTs is unknown to the BS and must be estimated by the RL algorithm. 
The only required property of the hash function is that it must uniformly map the inputs over the output range $\left\{1\dots L\right\}$, in order
to minimize slot selection collision. 
This can be achieved by means of simple functions that are well suited for execution at IoT devices with low computation power through the built-in MCU at each device. 
%green 

%probability \footnote{According to 3GPP TS38.300~\cite[Clause 16.10.6]{TS38300}, a device that remains in the \emph{inactive} state (i.e., without a packet in its queue) can also receive broadcast messages from the BS. This ability applies to ADDs in this study for their next attempt.}

%----------------------------------------------------------------------- 
\subsection{SDP Transmission Schemes}
\label{subsec:eval_schemes}
%----------------------------------------------------------------------- 
%{\color{red}This subsection was VI-B) in the initial submitted. To be integrated with III-B in the revised version.}
%
To facilitate random access and SDP data transmission in the envisaged network scenario, two schemes have been defined in our framework, as presented below. While Scheme A spreads the transmissions of SDPs %intra- and inter-cluster interference 
across slots in a frame with equal probability, Scheme B focuses on diminishing slot selection collisions 
%intra-cluster interference 
among $C_1$ devices. %than in $C_2$ }
%A summary of these schemes and their main features can be found in Table~\ref{tab:access_schemes}. %Note that constraining the number of devices that transmit per frame through access control leads to much lower inter- and intra-cluster interference than when no access control is exercised.
%-----------------------------------------------------------------------------
\subsubsection{Scheme A}
%-----------------------------------------------------------------------------
Devices in both clusters, $C_1$ and $C_2$, deploy the CB mode. 
However, different access probabilities might be assigned to each of them by the PG-driven access control algorithm. 
As $C_2$ devices are located farther away from the BS,
%when reward function $r^{(2)}$ is employed, 
a higher access probability is \emph{typically} assigned to $C_2$ 
devices to improve cluster throughput fairness among $C_1$ and $C_2$ devices.
%-----------------------------------------------------------------------------
\subsubsection{Scheme B}
%-----------------------------------------------------------------------------
While devices from $C_2$ still follow the CB mode, $C_1$ devices deploy the SCF mode. 
With the SCF mode, $C_1$ devices perceive much lower intra-cluster interference, leading to a higher successful SDP detection rate for $C_1$ transmissions. 
Devices from $C_2$ may also perceive a reduction in inter-cluster interference.
%may marginally perceive lower inter-cluster interference. Furthermore, note that both intra- and inter-cluster interference still exists, as multiple SDP transmissions from the same or different clusters may also occur simultaneously on the same radio resource.

%{\color{red}To be integrated with schemes.}
%One of the main motivations that lead to the proposal of the SCFAS is the fact that intra-cluster interference is dominant in the scenario of study, as observed in our previous studies~\cite{IoTJ23}\cite{PIMRC24}. Then, as will be shown later, the more effective deploying scenario is when the SCFAM is associated to devices of $C_1$, while the CBAM is associated to devices of $C_2$.  When the SCFAM is deployed by both, devices of $C_1$ and $C_2$, $C_2$ devices only perceive a marginal performance improvement. 

%----------------------------------------------------------------------- 
 \subsection{Performance Metrics}
\label{subsec:metrics}
%------------------------------------------------------------
The following four performance metrics are defined. 
\begin{itemize}
    \item \emph{Cluster throughput ($\gamma_i$ for cluster $C_i$}) and \emph{system throughput ($\gamma_s$)}.  
				They are defined as the average number of packets successfully transmitted per frame by a cluster and by the entire network including all clusters, respectively. 
				%system throughput specifies the total number of packets successfully transmitted per frame in the entire network, i.e., including all clusters. }
    
    \item \emph{Access delay (D)}. It is defined as the average number of frames it takes for a device to transmit a packet successfully. 
    This metric encompasses not only the frames during which an ADD defers its transmission 
		but also those frames where an SDP was transmitted but the BS failed to detect it.
		
		\item \emph{Device energy consumption (E)}. It is defined as the average energy consumed by a device 
			per successfully transmitted SDP.
			It aggregates the energy consumed by an active device: 
				i) to transmit an SDP; 
				ii) during retransmission(s); 
			 iii) for an ACK message reception upon a successful transmission; and 
			  iv) while it defers its transmission. %is not qualified for transmission.  

		\item \emph{Throughput fairness index $\widehat{J}(\gamma_1, \dots,\gamma_C)$}. It is the Jain's fairness index computed as 
		$\widehat{J}(\gamma_1, \dots,\gamma_C) = {(\sum_{i=1}^{C} \gamma_i)^2}/{(C \times \sum_{i=1}^{C} \gamma_i^2})$, 
         %  $\widehat{J}(\gamma_1, \gamma_2) = \left(\gamma_1 + \gamma_2\right)^2 / \left( C\cdot\left(\gamma_1^2 + \gamma_2^2\right)\right)$, 
			i.e., computed using the \textit{average} throughput, instead of the instantaneous throughput as defined in~(\ref{fairness_index}).  
			$\widehat{J}(\gamma_1, \dots, \gamma_C) \in \left[1/C, 1\right]$. 
			Then, the closer the $\widehat{J}(\gamma_1, \dots, \gamma_C)$ to $1$, 
			the fairer the throughput distribution among clusters. %{\color{red}I think it is better to keep it as general, for C clusters. } 
%compares the throughput obtained by two or multiple clusters, 
\end{itemize} 
%------------------------------------------------------------

%----------------------------------------------------------------------- 
%\subsubsection{Energy Consumption}
%\label{subsec:transmission_principle}
%----------------------------------------------------------------------- 

%===============================================================
%\section{Framework PG ALGORITHMS} %\vspace{-2mm}
%\label{sec:policylearning} 
%\label{Sec:Scheme}
%\section{{\color{blue}PG Algorithm for Access Control}} %\vspace{-2mm}
\section{Policy Gradient Supporting the Framework}
\label{sec:RL-agent} 

%===============================================================
%In this section we define the policy gradient algorithms that support the operation of the proposed framework~\cite{RL_book}. 

%
%%%%%%%%%%%%%%%%%%%%%%%%%%%%%%
\begin{figure}
	\centering
    \includegraphics[scale = 0.58]{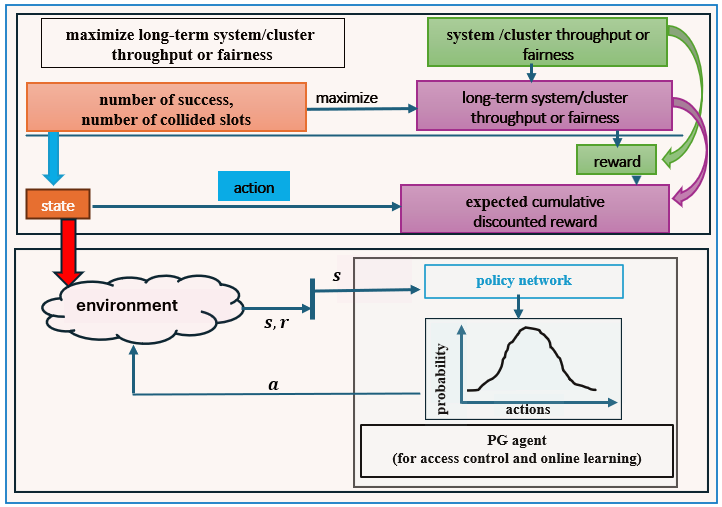} %\vspace{-4mm}
\caption{PG-driven online learning: Access control and seed generation.} %\vspace{-5mm}
	\label{fig:RL_algorithms}
\end{figure} 
\vspace{-1mm}
PG algorithms learn parametrized policies that can select actions without consulting an action-value function~\cite{RL_book}.
The policies learned through PG are parametrized functions defining the probability of each action. 
%{\color{blue}The PG-based access control mechanism presented in this section constitutes a core component} 
In the RL4SCF framework, the PG algorithm learns two action policies for each cluster $C_i$: one for access probability $a_i$, and the other for seed $b_i$. 
Both policies are learned simultaneously by the same RL-agent.
%-----------------------------------------------------------------
\vspace{-1mm}
\subsection{States and Actions}
%-----------------------------------------------------------------
We define the system state $s$ as the total number of successful transmissions observed by the BS in the preceding frame, including the successful transmissions from all clusters.
Let $s_i$ denote the number of successful transmissions in cluster $C_i$ in the previous frame. Then $s=\sum_{i=1}^C s_i$. 
The state space is discrete, $ s \in \{ 0,\dots, \sum_{i=1}^C N_i \} $. %where $N_i$ is the number on devices in cluster $C_i$.}
%{\color{blue}as $s \in \{0, \dots, \sum_{i=1}^C s_i\}$, with $n$ elements}. {\color{red}Check.}

The action is a combination of a vector of access probabilities with an element $a_i$ for every cluster $i$, 
and a vector of seeds, with an element $b_i$ for every cluster operating in the SCF mode.
The access probabilities are continuous within the interval of $[0.1,1]$, 
while the seeds are taken from a discrete set of candidate seeds.

%-----------------------------------------------------------------
\vspace{-1mm}
\subsection{Policy Learning}
%-----------------------------------------------------------------
Let $\pi_i(\cdot|s, \vec{\theta}_i)$, with parameters vector $\vec{\theta}_i$, 
be the access probability policy for cluster $C_i$,
with $\pi_i(a_i|s,\vec{\theta}_i)$ being the probability of access probability $a_i$ when the system state is $s$. 
Analogously, let $\tau_i(\cdot|s,\vec{\phi}_i)$ be the seed policy for cluster $C_i$,
with parameters $\vec{\phi}_i$.
These policies are learned through updating, at every step, 
the parameters vectors:
\begin{eqnarray}\label{eq:update1}
\vec{\theta}_i 
& \longleftarrow  & \vec{\theta}_i+\alpha_{\theta}\; \delta\; \nabla\log\pi_i(a_i|s,\vec{\theta}_i)	\; , \\ \label{eq:update2}
\vec{\phi}_i & \longleftarrow   & \vec{\phi}_i+\alpha_{\phi}\; \delta\; \nabla\; \log\tau_i(b_i|s,\vec{\phi}_i)	\; ,
\end{eqnarray} 
where $\alpha_{\theta}$ and $\alpha_{\phi}$ are learning steps.
The error $\delta$ is computed from a learned state-value function $V(s,\vec{\omega})$, 
\begin{equation}%\nonumber
\delta=  r + \epsilon \; V(s_{\text{next}},\vec{\omega})-V(s,\vec{\omega})	\; ,
\end{equation} 
where $s$ is the current state, $s_{\text{next}}$ the next state, $r$ the reward and $\epsilon$ a discount rate.
The state-value function is linearly approximated by $V(s,\vec{\omega})=\omega^{(s)}$, 
where $\omega^{(s)}$ is the $s$-th element of the $n$-dimension vector $\vec{\omega}$, 
and is updated by stochastic gradient descent optimization. 
As all the elements of the gradient are $0$  except the $s$-th element, which is $1$,
at state $s$  only the $s$-th element of $\vec{\omega}$ is updated:
\begin{equation}%\nonumber
\omega^{(s)} \longleftarrow \omega^{(s)} + \alpha_{\omega}\; \delta	\; ,
\end{equation}
where $\alpha_{\omega}$ is a learning step. 
Policies $\pi_i(\cdot|s,\vec{\theta}_i)$ are continuous probability distributions within the range of $[0.1,1]$.
For them, we use log-normal distributions $\log a'_i\backsim \mathcal{N}(\mu_i,\sigma^2)$ transformed as 
\begin{equation}%\nonumber
a_i=\frac{0.1+a'_i}{1+a'_i}	\; .
\end{equation}
A single value of the variance $\sigma^2$ is fixed to tune the dispersion of the distributions,
while each mean $\mu_i$ is linearly approximated by  
$\mu_i(s,\vec{\theta}_i)=\theta_i^{(s)}$, where $\theta_i^{(s)}$ is the $s$-th element of the $n$-dimension vector $\vec{\theta}_i$.
From~\eqref{eq:update1},
it follows that, for state $s$ and action $a_i$, only the $s$-th element of $\vec{\theta}_i$ is updated:
\begin{equation} %\nonumber
\theta^{(s)}_i \longleftarrow  \theta^{(s)}_i +\alpha_{\theta}\; \delta\; \frac{\log \frac{a_i-0.1}{1-a_i}-\theta_i^{(s)}}{\sigma^2}	\; .
\end{equation} 
Policies $\tau_i(\cdot|s,\vec{\phi}_i)$ are discrete probability distributions with a value at state $s$ for each candidate seed.
We apply soft-max distributions,
\begin{equation}%\nonumber
\tau_i(b^{(j)}|s,\vec{\phi}_i)=\frac{e^{h_i(s,b^{(j)},\vec{\phi}_i)}}{\sum_{b}e^{h_i(s,b,\vec{\phi}_i)}},
\end{equation}
where $h_i(s,b^{(j)},\vec{\phi}_i)=\phi_i^{(s,j)}$ are parameterized
with $n\times q$-dimension vectors $\vec{\phi}_i$,
with $q$ equal to the number of candidate seeds. For state $s$ and seed $b_i$, 
every element $(s,j)$ of $\vec{\phi}_i$, for $j=1\dots q$, is updated as 
\begin{equation} %\nonumber
\phi^{(s,j)}_i  \longleftarrow 
\begin{cases}
 \phi^{(s,j)}_i+\alpha_{\phi}\; \delta -\tau_i(b^{(j)}|s,\vec{\phi}_i) &  \text{if } b^{(j)}= b_i \\ %\nonumber
  \phi^{(s,j)}_i -\tau_i(b^{(j)}|s,\vec{\phi}_i) & \text{if } b^{(j)}\neq b_i	\; .
\end{cases}
\end{equation}

%-----------------------------------------------------------------
\vspace{-6mm}
\subsection{Rewards}
%-----------------------------------------------------------------
Two reward functions are defined. While the goal of the first reward function is to maximize the number of packets that can be successfully transmitted per frame,
the second one intends to achieve certain degree of fairness among clusters.

Reward function $r^{(1)}$ is the total number of successful transmissions in the previous frame:
\begin{equation}
   r^{(1)}(s_1, \dots,s_C) = \sum_{i=1}^{C} s_i =s \; .
   \label{reward 1}
\end{equation}

As an alternative to promote fairness in terms of throughput among clusters, we define reward function  $r^{(2)}$ as 
\begin{equation}
   r^{(2)}(s_1, \dots,s_C) = r^{(1)}(s_1, ...,s_C)\; J(s_1, \dots,s_C)\,,
   \label{reward_2}
\end{equation}
where $J(s_1, \dots,s_C)$ is the instantaneous Jain's fairness index obtained solely based on the observed number of successes \emph{in the preceding frame}, 

%where $J$ is the  Jain's Fairness Index, given by
\begin{equation}
    J(s_1, \dots,s_C) = \frac{(\sum_{i=1}^{C} s_i)^2}{C \sum_{i=1}^{C} s_i^2}\; .
    \label{fairness_index}
\end{equation}
%===========================================================================================

% -----------------------------------------------------------------------
\begin{algorithm}[tbp]\small
\caption{Implementation of the PG Algorithm}
\label{alg:PGradient}
\DontPrintSemicolon
Set learning steps $\alpha_{\theta}, \alpha_{\phi}, \alpha_{\omega} $  \\
\For{each  cluster $i$}{
Initialize access probability policy $\pi_i$: $\vec{\theta}_i\leftarrow\vec{0}$\\
Initialize seed policy $\tau_i$: $\vec{\phi}_i\leftarrow\vec{0}$}
Initialize state-value function $V$: $\vec{\omega}\leftarrow\vec{0}$   \\
Initialize state: $s\leftarrow 0$ \\
 \For{each step}{
 	\For{each  cluster $i$}{
  	Choose access probability $a_i\backsim \pi_i(\cdot|s,\vec{\theta}_i)$ \\
  	Choose seed $b_i\backsim \tau_i(\cdot|s,\vec{\phi}_i)$  \\
    Broadcast $a_i$ and $b_i$}
  Transmit and observe $successes$\\
    Set next state: $s_{\text{next}}$=  $successes$\\
    Compute reward: $r=reward(successes)$\\
	Compute error:  $\delta= r +\epsilon\; V(s_{\text{next}},\vec{\omega})-V(s,\vec{\omega})$  \\
	\For{each  cluster $i$}{
	Update  $\pi_i$: $\vec{\theta}_i \leftarrow \vec{\theta}_i + \alpha_{\theta}\; \delta\; \nabla\log\pi_i(a_i|s,
	\vec{\theta}_i)$ \\
	Update  $\tau_i$: $\vec{\phi}_i \leftarrow \vec{\phi}_i + \alpha_{\phi}\; \delta\; \nabla\log\tau_i(b_i|s,
	\vec{\phi}_i)$ 
}
	Update  $V$:	$\vec{\omega} \leftarrow \vec{\omega} + \alpha_{\omega}\; \delta\; V(s,\vec{\omega})$ \\
	$s\leftarrow s_{\text{next}}$ 
} 
\end{algorithm} %\vspace{-3mm}
% ------------------------------------------------------------------------
%

%------------------------------------------------------------
\begin{table}[!t] \begin{center}\small
	\caption{PG hyper-parameters} 
	\label{table:PGparameters} 
	\centering 
	\scriptsize
	\begin{tabular}{ |l |l| l| }\toprule
		{\it Parameter} & Symbol & {\it Value }  \\ \toprule  
		Learning step for state-value function  & $\alpha_{\omega}$ & 0.001\\
		Learning step for $a_i$ policy & $\alpha_{\theta}$ & 0.001\\
		Learning step for $b_i$ policy  & $\alpha_{\phi}$ & 0.01\\
		%State-value parameters vector dimension & $ n $ & number of devices +1\\
		Number of states & $ n $ & number of total devices +1 \\
		%$b_i$ policy parameters vector dimension & $ q $ &  10 \\
		Number of candidate seeds  & $ q $ &  10\\
     %   $a_i$ policy parameters vector dimension & $ n $ & number of devices +1 \\
		%$b_i$ policy parameters vector dimension & $ q $ &  10 \\
		$a_i$ policy standard deviation & $\sigma$ & 0.1\\
		Discount rate & $\epsilon$ & 0.5 \\	
				\bottomrule
	\end{tabular}
	\end{center} \vspace{-9mm}
\end{table}
%------------------------------------------------------------
%

%===============================================================
%\section{SIMULATION MODEL SETUP} %\vspace{-2mm}
%\label{Sec:Scheme}
%===============================================================

\section{Implementation Overview} %\vspace{-2mm}
\label{sec:implementation}
In this section, we first summarize the implementation
of the RL-enabled access control mechanism at the BS and slot selection through hashing by devices and then
explain how energy consumption for a device is calculated.

%----------------------------------------------------------------------- 
\subsection{Implementation of the PG Algorithm}
Alg.~\ref{alg:PGradient} illustrates how PG-based access control is implemented. 
This algorithm is continuously running at the BS and taking actions in real-time, in a frame-by-frame manner or at a configurable update interval that may cover multiple frames. In Table~\ref{table:PGparameters}, we present a list of hyper-parameters and their values that are adopted in our implementation.

%----------------------------------------------------------------------- 
\subsection{Implementation of the Hash Function}
%\label{subsec:transmission_principle}
%-----------------------------------------------------------------------
The hash function~(\ref{eq:hashF}) adopted by the devices has been implemented using the random number generator $\mathrm{rand}()$, which is a function that generates \emph{pseudo-random integers} with minimal computational cost and this function is available in any programming language. Each device at cluster $C_i$ calculates the slot as 
	\begin{eqnarray} 
		\mathrm{srand}(b_i + \text{ID})\,, 
		\label{hash_a}\\
		slot= \mathrm{rand}(L),
		\label{hash_b}
	\end{eqnarray} 
where, as previously defined, $b_i$ is the hash seed broadcasted by the BS to cluster $C_i$, ID is a unique identifier of the device, and
$L$ is the number of time slots in one frame.

To compute the time slot assignment for a device, its initial state is set to be $(b_i+\text{ID})$ and the slot assigned to the device is obtained by  
$rand()$\text{ mod} $L$.
This time slot assignment procedure does not lead to random mapping but achieves ultimately a hash table that maps devices to time slots within the interval of $\left[ 1, L\right]$, with the lowest number of collisions for the current set of transmitting devices.

%$\mathrm{rand}(L)$ generates a random integer between $1$ and $L$, and $\mathrm{srand}()$ is the function that sets the starting point of the pseudo-random sequence of integers generated by $\mathrm{rand}()$.     
%the result of executing operations~(\ref{hash_a}) and~(\ref{hash_b}) to the IDs of the active devices in the cluster, 

%To assess the operability and performance of the PG-driven access control mechanism, we have tested the algorithm with the simulation setup described in Subsec.~\ref{sec:netconf}.

\begin{table}[!t] \small
	\caption{Parameters for energy consumption calculation~\cite{nRF9160}} \label{table:EnergyParameter} 
	\vspace{-1mm}
	\centering
	%\resizebox{0.9\columnwidth}{!}{%       
	\scriptsize
	\begin{tabular}{ |l| l| l || l| l| l| }\toprule
		{\it Parameter} & {\it Value } & \it Unit & {\it Parameter} & {\it Value} & {\it Unit} \\ \toprule  %\cline{1-4}
	%	\multirow{1}{*}{Common} & & & Supply voltage & 3 & V \\ \midrule
		%\multirow{8}{*} {WuR device} & & & Data rate & 250 & kbps \\ %\cline{2-4}
	Slot duration &  20  & ms    & Transmit power    & 200 & mW \\ %\cline{2-4}
		Packet size   & 128  & bytes & Reception current &  35 & mA \\ %\cline{2-4}
		ACK size      &  16  & bytes & Idle current &  2.7 & $\mu$A  \\ %\cline{2-4}
		Data rate     &  60  & kbps  & Voltage      &  3.7 & volt   \\ %\cline{2-4}
		\bottomrule
	\end{tabular}
	\vspace{-5mm}
\end{table}
%--------------

%----------------------------------------------------------------------- 
\vspace{-3mm}
\subsection{Device Energy Consumption Calculation}
%----------------------------------------------------------------------- 
%For device energy consumption, we calculate the average amount of energy a device needs to consume in order to transmit a data packet successfully. 
We consider that a device may transmit several times across multiple frames before its transmission is acknowledged by the BS as successful through an ACK message. For a successful transmission, both SDP transmission and ACK reception occur inside one slot~\cite{5G_book}.
%\cite[Section 7.4.4.3.1]{TR45820}}. 

Denote respectively by $P_{tx}$ and $P_{rx}$ the transmission and the reception power for a device, 
$T_{tx}$ the transmission time for one SDP, 
%{\color{red}\textbf{The time it takes to tx a single SDP should be one time slot.} \textit{Not necessarily - it depends on packet size and transmission data rate. The values we used are listed in Tab. II.}}
$T_{rx}$ the reception time for ACK, 
$T_{slot}$ the slot duration, 
%$L$ the number of slots in frame, 
$N_{tot}$ the total number of attempts for one successful transmission among which $N_{tot}-1$ failed and one succeeded, 
and $N_{idle}$ the number of frames that an ADD defers its transmission. 
Then the total energy consumed by a device per successfully transmitted SDP, $E_{tot}$, is calculated as follows. 
%\vspace{-3mm}
\begin{align}
    E_{tot} =  P_{tx} T_{tx} N_{tot}  +  P_{rx} T_{rx} + P_{idle} T_{slot} L N_{idle}. 
\end{align}
In the above expression, which does not rely on any specific type of IoT devices, we assume that the energy consumed during %the \emph{loop back delay} which is 
the interval between a successful data packet transmission and its ACK message reception inside the same time slot %and for the overhead of protocol handshake 
is negligible. Furthermore, the power consumed by a device in the idle state, $P_{idle}$, is typically three orders of magnitude lower than the power consumed while transmitting and receiving.

As a realistic example, we provide in Table \ref{table:EnergyParameter} a list of parameters and values that are adopted in our energy consumption calculations \cite{nRF9160}. In this example, the slot duration is configured as 20 ms as it requires (128 + 16) bytes$\times$8/60 kbps = 19.2 ms to transmit an SDP and receive its corresponding ACK message upon a successful transmission. Note however that these parameters are configurable and the operation of our schemes is irrelevant to the slot or frame duration. 
%and most of these values are collected from the Nordic Semiconductor \emph{nRF9160 product specification} for narrowband IoT (NB-IoT) . %\footnote{{\color{red}For the sake of simplicity, the slot duration and frame structure are specifically defined for this study. For our energy consumption calculation, we do not intend to follow exactly the NB-IoT specifications for all parameters.}}
The numerical results reported in Section \ref{sec:results} are the mean values of per device energy consumption averaged over all successfully transmitted SDPs. 
%packets that are transmitted during the simulation duration.   
%

%===============================================================
%\section{NUMERICAL RESULTS AND DISCUSSIONS} %\vspace{-2mm}
%\label{Sec:Scheme}
%===============================================================

\section{Numerical Results and Discussions} %\vspace{-2mm}
\label{sec:results}
In this section, we first explain the different network configurations %deployed and the two evaluation schemes. 
that have been designed to assess the benefits that the RL4SCF framework can bring on network performance. 
Then, we present and discuss the performance evaluation results that are obtained through extensive simulations.  

\subsection{Network Configuration}
\label{subsec:network_configuration}
The RL4SCF framework illustrated in Fig. \ref{fig:framework} has been implemented based on a custom-built simulator we have developed using Java. The implemented network supports a single-cell network with a number of static devices uniformly distributed across two distinct clusters.
%A custom-built simulation model of the RL4SCF framework has been implemented in Java.   

To evaluate the performance of the framework, we assume that packet arrivals to devices follow a Bernoulli distribution with an arrival probability \emph{per frame} $\lambda \in [0.1,1]$. Note that any other arrival distribution or arrival pattern may also be applied for performance evaluation.

%Packet arrivals to devices in all clusters follow a Bernoulli distribution with arrival probability $\lambda \in [0.1,1]$. {\color{blue}The Bernoulli distribution has been adopted for the numerical results presented below, but our framework can also work with any other arrival distribution or arrival pattern.}
%However, the proposed RL4SCF framework supports also other arrival distribution or arrival pattern.} 

We define different network configurations, where the number of devices per cluster, $N_1$ or $N_2$, 
increases as the number of time slots per frame $L$ becomes larger. For the numerical results reported in this paper, the following configuration parameters have been adopted, namely, $L = \{4, 8, 16\}$ and $N_1 = N_2 =\{8, 16, 32\}$ respectively. 
%and {\color{blue} \sout{$N= $}}.
For simplicity, we refer to a specific network configuration by a tuple $\left\{L;N_1+N_2\right\}$.
For a complete definition of the physical layer and network configuration parameters adopted along the performance evaluation process, 
please refer to Table \ref{table:NetworkConfiguration}.

%--------------------------------------------------------------------------------------------------------
\begin{table}[!t] \small
	\caption{Physical layer and network configuration} \label{table:NetworkConfiguration} 
	%\vspace{-1mm}
	\centering
	%\resizebox{0.9\columnwidth}{!}{%       
	\scriptsize
	\begin{tabular}{ |p{24mm} |p{7mm} || p{27mm} |p{12mm}| }\toprule
		{\it Parameter} & {\it Value } & {\it Parameter} & {\it Value} \\ \toprule   
     Number of clusters ($C$)   &  2    &  Standard deviation for shadow fading & 8 dB  \\ 
     C1 to BS distance ($d_1$)  & 450 m  & Receiver sensitivity  & -104 dBm      \\
    C2 to BS distance ($d_2$)	&  900 m   & No. of devices/cluster ($N$)  &  \{8; 16; 32\} \\
		Cluster radius     &  25 m    & No. of slots/frame ($L$) & \{4; 8; 16\} \\ 
    SIC SINR threshold ($\beta$) & 10 dB & Arrival probability ($\lambda$) & [0.1,...,1] \\
    Antenna height ($g$) & 30 m  & Noise power spectral density  ($\eta^2$)  & -174 dBm/Hz \\    
		\bottomrule
	\end{tabular}
	\vspace{-3mm}
\end{table}
% ----------------------------------------------------------------------------
%	

% -------------------------------------------------------------------------------------
\begin{figure*}[ht]
  \centering
  \begin{subfigure}{0.32\textwidth}\vspace{1mm}
        \includegraphics[scale = 0.4]{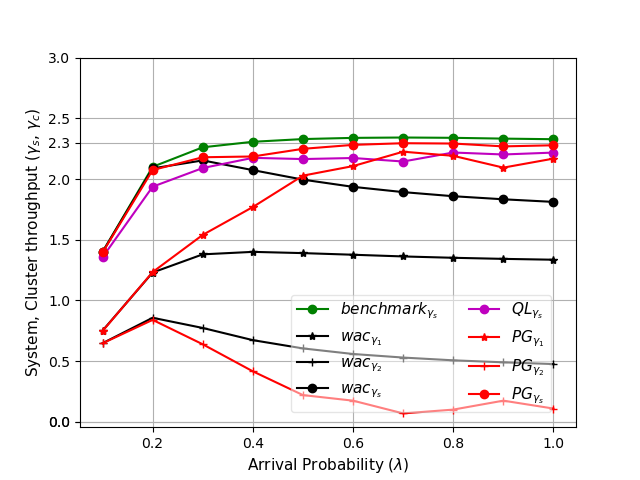}\vspace{-2mm}
        \caption{$\{4; 8+8\}$.} \vspace{-1mm}
        \label{DQN_based_Thp_884}
  \end{subfigure}
  \begin{subfigure}{0.32\textwidth}\vspace{1mm}
        \includegraphics[scale = 0.4]{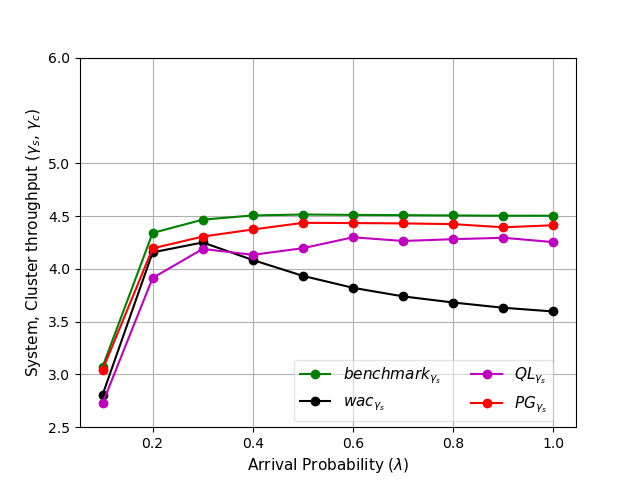}\vspace{-2mm}
        \caption{$\{8;16+16\}$.} \vspace{-1mm}
        \label{DQN_based_Thp_16168}       
  \end{subfigure} 
  \begin{subfigure}{0.32\textwidth}
        \vspace{-1mm}
        \includegraphics[scale = 0.4]{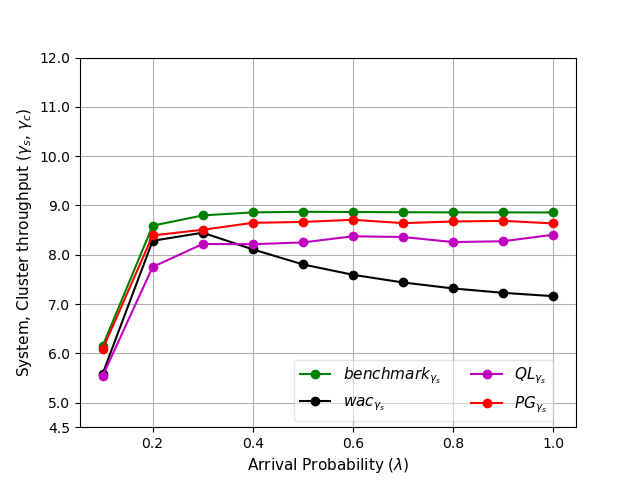}\vspace{-2mm}
        \caption{$\{16;32+32\}$.} \vspace{-1mm}
        \label{DQN_based_Thp_323216}      
  \end{subfigure} 
  \caption{Throughput in Scheme A when reward function $r^{(1)}$ is adopted.}  
  \label{fig:schemeA_reward1}
\vspace{-6mm}
\end{figure*}
%%%%%%%%%%%%%%%%%%%%%%%%%%%%%%
\begin{figure*}[ht]
  \centering
  \begin{subfigure}{0.32\textwidth}\vspace{1mm}
        \includegraphics[scale = 0.4]{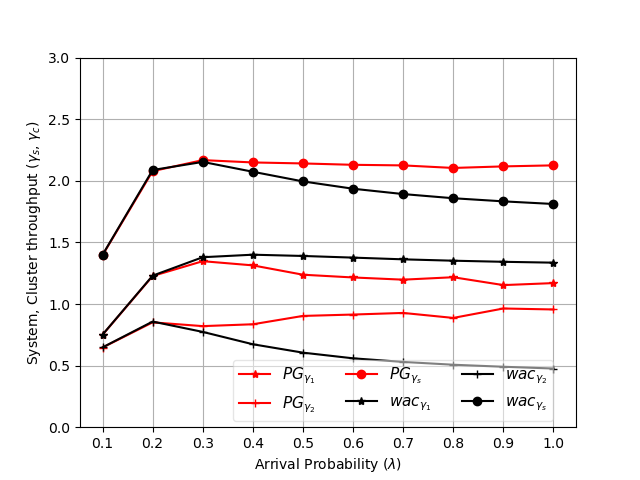} \vspace{-6mm}
        \caption{$\{4;8+8\}$.} \vspace{-2mm}
	\label{Fair_Thp_8_8_4}
  \end{subfigure}
  \begin{subfigure}{0.32\textwidth}\vspace{1mm}
        \includegraphics[scale = 0.4]{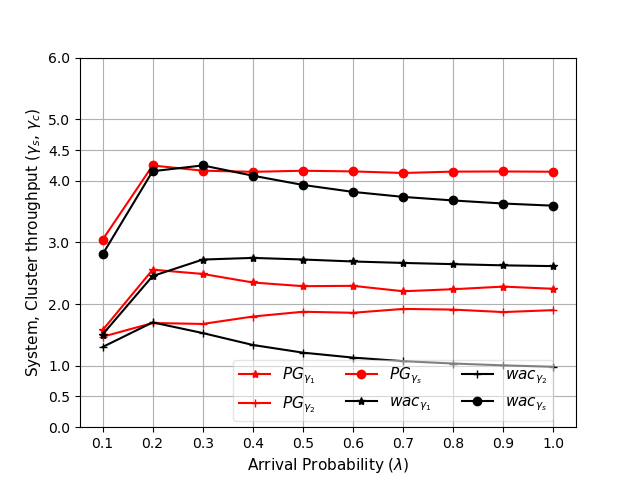} \vspace{-6mm}
        \caption{$\{8;16+16\}$.} \vspace{-2mm}
	\label{Fair_Thp_16_16_8}       
  \end{subfigure} 
  \begin{subfigure}{0.32\textwidth}
        \vspace{-1mm}
        \includegraphics[scale = 0.4]{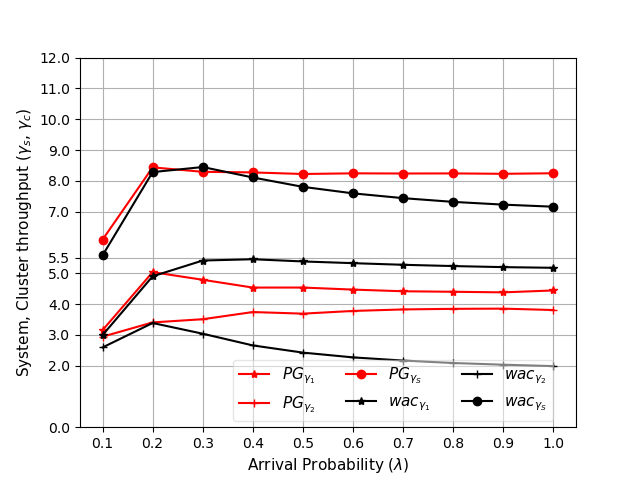} \vspace{-6mm}
        \caption{$\{16;32+32\}$. } \vspace{-2mm}
	\label{Fair_Thp_32_32_16}      
  \end{subfigure} 
  \caption{Throughput in Scheme A when reward function $r^{(2)}$ is adopted.} 
  \label{fig:schemeA_reward2}
\vspace{-6mm}
\end{figure*}
%=====================================================================================================

%-----------------------------------------------------------------------------
\subsection{System and Cluster Throughput in Scheme A}
%-----------------------------------------------------------------------------
In Scheme A, with both clusters operated in the CB mode, we evaluate:  
1) the benefit of introducing access control;   
2) how close our access control mechanism can approach an upper-bound throughput benchmark; 
3) scalability of the framework; and 
4) the impact that the reward function has on the pursued performance objectives.     
%-----------------------------------------------------------------------------
\subsubsection{System throughput with and without access control} 
%-----------------------------------------------------------------------------
For comparison purposes, we illustrate in Fig.~\ref{fig:schemeA_reward1}  the throughput achieved both by Scheme A and another reference scheme, referred to as \textit{without access control} ($wac$), in which the access control mechanism in RL4SCF 
%designed for the RL4SCF access control framework '
is \emph{disabled}. For $wac$, neither access control nor CB or/and SCF modes are introduced.   
%i.e., without enabling the broadcast of per cluster access probabilities to support both the CB and SCF modes, and hash seed to support the SCF mode. We refer to the obtained throughput in this case %lower-bound benchmark} as the throughput \textit{without access control} ($wac$).

Fig.~\ref{fig:schemeA_reward1} depicts the variation of the obtained system throughput ($\gamma_s$) and cluster throughput ($\gamma_i$) with the traffic load ($\lambda$). 
%It compares the throughput achieved when devices deploy the CB mode supported by the broadcast of access probabilities from the BS, and the one achieved by \textit{wac}. %when devices contend without access control (\textit{wac}) being exercised by the BS. 
For the purpose of comparison, a curve (in violet) representing the system throughput under the same network configuration obtained 
from our earlier work~\cite{PIMRC24} which performed a QL-based access scheme is also kept in this figure.  

Before the saturation point, the achieved throughput increases with the traffic load for all the evaluated schemes. %algorithms. 
Without access control, a significant decline in system and cluster throughput is observed when $\lambda > 0.3$. 
Clearly, the benefit of applying access control becomes evident when the traffic load grows to a certain level ($\lambda > 0.3$), 
as observed in Fig.~\ref{fig:schemeA_reward1} for all the three network configurations.
%-----------------------------------------------------------------------------
\subsubsection{System throughput versus upper-bound benchmark} 
%-----------------------------------------------------------------------------
%As a \emph{system throughput upper-bound benchmark}, we also obtain an the system throughput computed 
The throughput upper-bound benchmark (the green curve in Fig.~\ref{fig:schemeA_reward1}) is obtained through exhaustive search of
the access probabilities that lead to the maximum system
throughput for each load level. %In other words, RL is not used to obtain it.
%In Fig.~\ref{fig:schemeA_reward1}, we present also the \textit{throughput benchmark} (a curve marked in green) which is obtained by \emph{exhaustive search} of the access probabilities that lead to the maximum system throughput for each load level without applying RL.  

Despite the fact that the access probability policies determined by the PG and QL algorithms are computed based 
on the partially observed system state information, both of them lead to
%the obtained policies achieve 
superb performance. 
In terms of system throughput, the ones achieved by the PG algorithm proposed in this work and the QL algorithm that was obtained from~\cite{PIMRC24} are quite close to the 
system throughput benchmark which is the performance upper-bound. %, i.e., the green curve in Fig.~\ref{fig:schemeA_reward1}.
 
%{\color{red} I do not think it is a good idea to confront the performance of PG and QL computing access probability policies. There possibly are an unlimited number of possible implementations of both algorithms. That is, another QL implementation might turn out to achieve better performance then the one achieved by the current PG algorithm.}

% -------------------------------------------------------------------------------------
\begin{figure*}[ht]
  \centering
  \begin{subfigure}{0.32\textwidth}\vspace{1mm}
        \includegraphics[scale = 0.4]{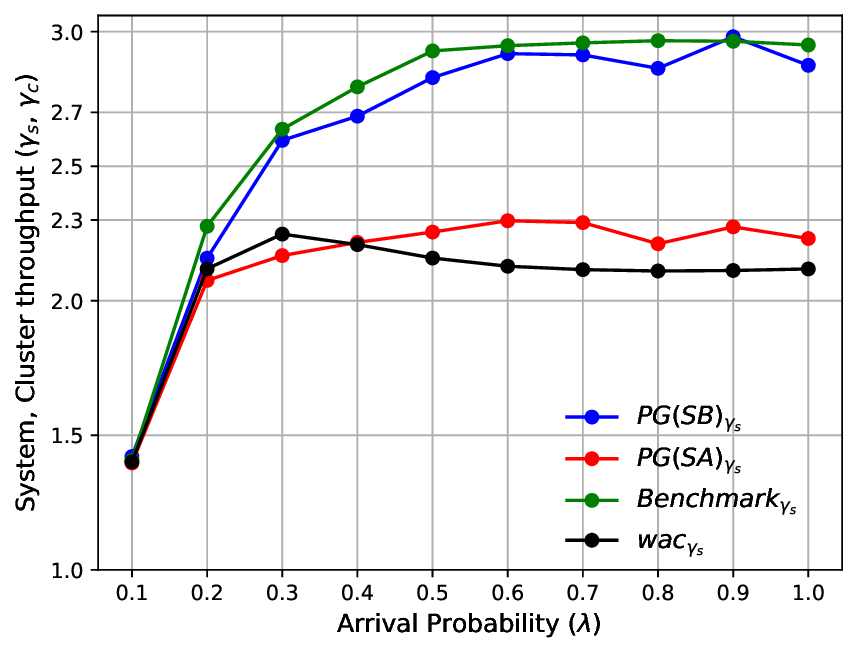} \vspace{-6mm}
        \caption{ $\{4;8+8\}$.} \vspace{-1mm}
	\label{Hash_C0_Thp_8_8_4}
  \end{subfigure}
  \begin{subfigure}{0.32\textwidth}\vspace{1mm}
        \includegraphics[scale = 0.4]{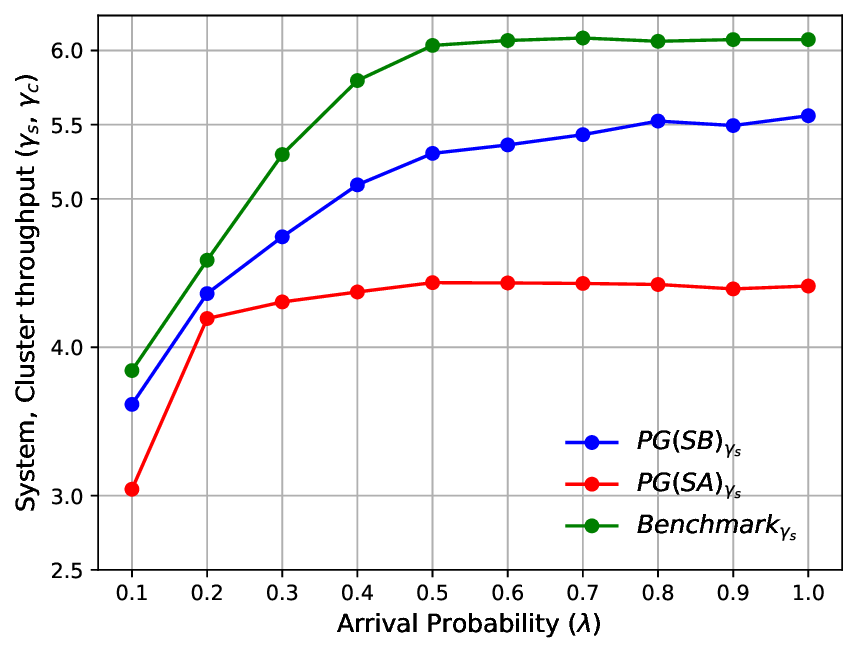} \vspace{-6mm}
\caption{$\{8;16+16\}$.} \vspace{-1mm}
	\label{Hash_C0_Thp_16_16_8}       
  \end{subfigure} 
  \begin{subfigure}{0.32\textwidth}
        \vspace{-1mm}
        \includegraphics[scale = 0.4]{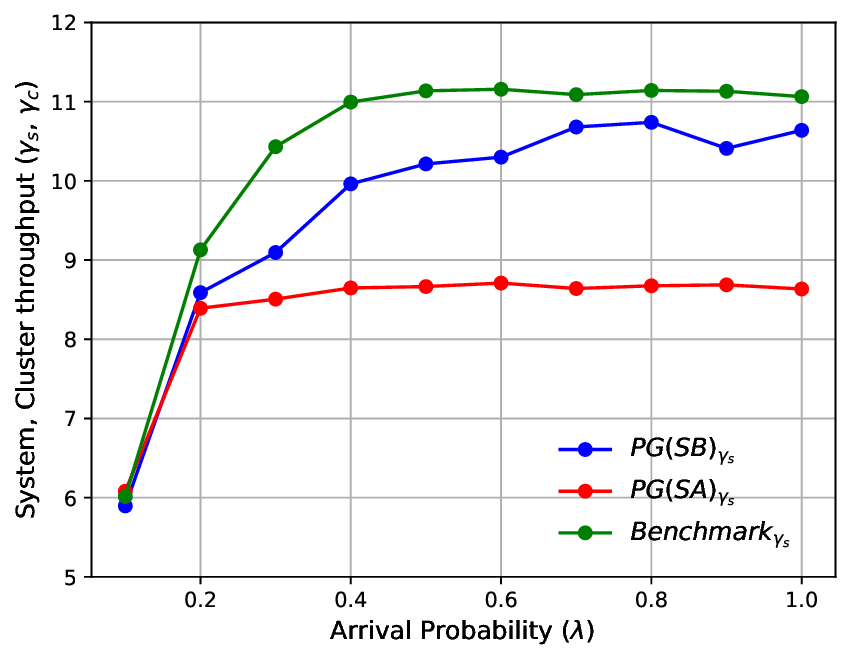} \vspace{-6mm}
    \caption{$\{16;32+32\}$.} \vspace{-1mm}
	\label{Hash_C0_Thp_32_32_16}    
  \end{subfigure} 
  \caption{Throughput in Scheme B when using reward function $r^{(1)}$.}  
  \label{fig:schemeB_reward1}
\vspace{-6mm}
\end{figure*}
% -------------------------------------------------------------------------------------
\begin{figure*}[ht]
  \centering
  \begin{subfigure}{0.32\textwidth}\vspace{1mm}
        \includegraphics[scale = 0.4]{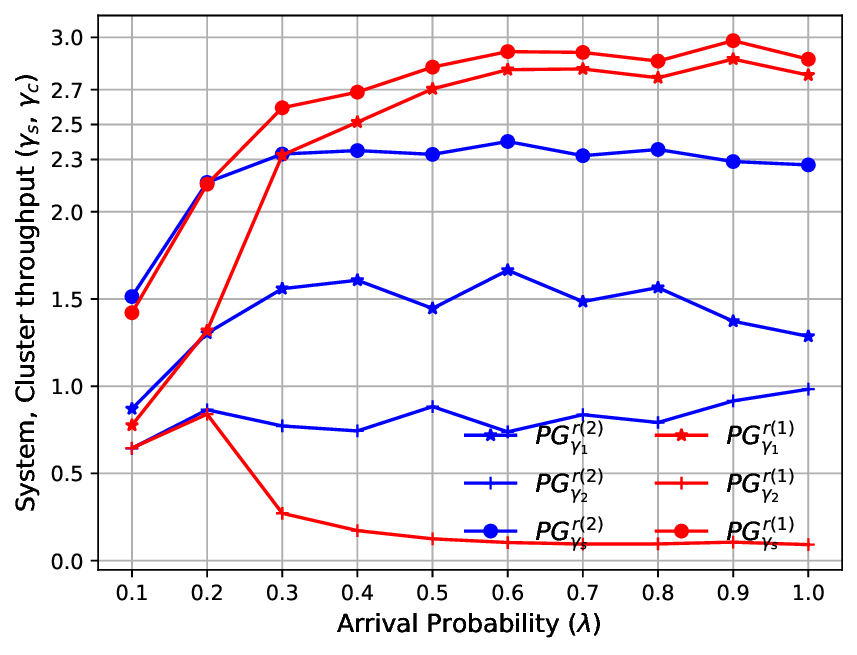} \vspace{-6mm}
        \caption{ $\{4;8+8\}$.} \vspace{-1mm}
	\label{Hash_Fair_C0_Thp_8_8_4}
  \end{subfigure}
  \begin{subfigure}{0.32\textwidth}\vspace{1mm}
       \includegraphics[scale = 0.4]{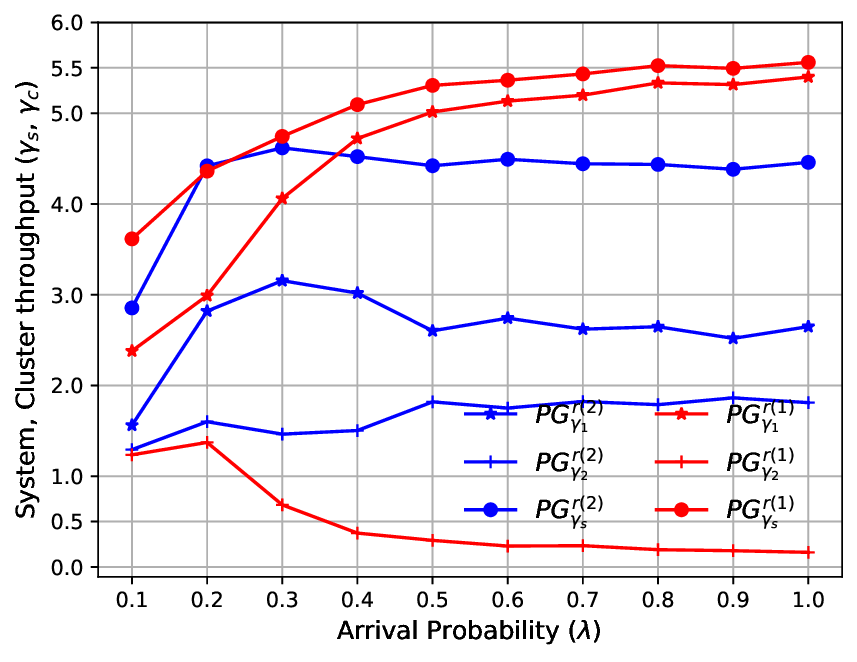} \vspace{-6mm}
\caption{ $\{8;16+16\}$.} \vspace{-1mm}
	\label{Hash_Fair_C0_Thp_16_16_8}     
  \end{subfigure} 
  \begin{subfigure}{0.32\textwidth}
        \vspace{-1mm}
        \centering
	\includegraphics[scale = 0.4]{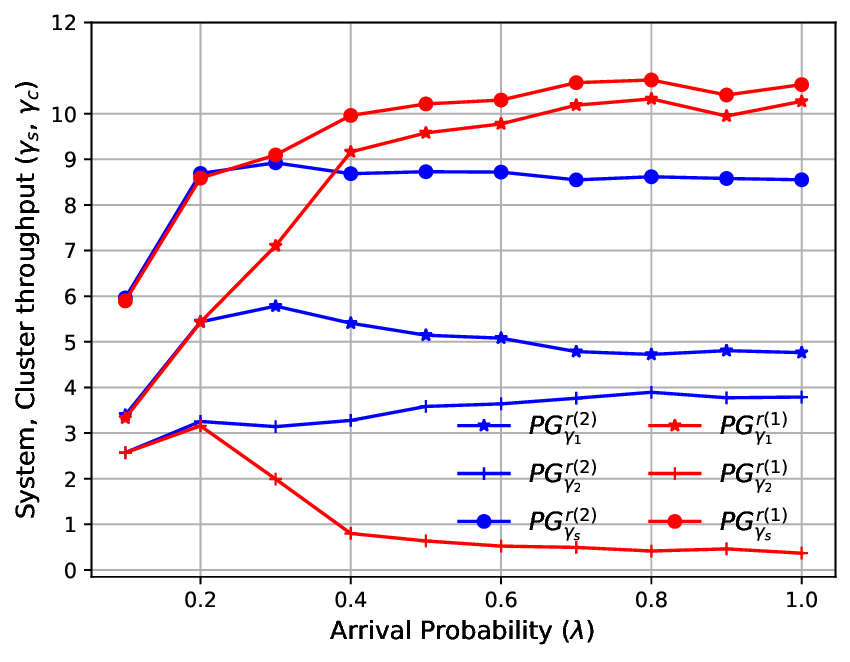} \vspace{-6mm}
        \caption{ $\{16;32+32\}$.} \vspace{-1mm}
	\label{Hash_Fair_C0_Thp_32_32_16}   
  \end{subfigure} 
  \caption{Throughput in Scheme B when using reward functions $r^{(1)}$ and $r^{(2)}$.} 
  \label{fig:schemeB_reward2}
\vspace{-7mm}
\end{figure*}

%-----------------------------------------------------------------------------
\subsubsection{Throughput fairness} 
%-----------------------------------------------------------------------------
In Fig.~\ref{fig:schemeA_reward2}, we focus on the throughput achieved by each cluster for the same network configurations 
studied above. 
In this figure, we also include the system throughput as a reference and refer to the cluster throughput achieved by RL4SCF
%the proposed RL4SCF access control framework 
as PG.
As it is evident, the difference between the cluster throughput of the two clusters becomes much smaller when reward function $r^{(2)}$ is adopted, 
compared to the difference between the ones shown in Fig.~\ref{fig:schemeA_reward1}a), where function $r^{(1)}$ was used.
%We denote by $\widehat{J}(\gamma_1, \gamma_2)$ the fairness index computed using the average throughputs. 
Correspondingly, the fairness indexes $\widehat{J}(\gamma_1, \gamma_2)$ for $\lambda = 1.0$ achieved in Fig.~\ref{fig:schemeA_reward2} 
are $0.998$, $0.996$, and $0.996$ for the three studied network configurations, 
\( \{4;8+8\} \), \( \{8;16+16\} \), and \( \{16;32+32\} \), respectively. 
These values indicate that with $r^{(2)}$ near-perfect throughput fairness between the two clusters has been achieved, 
demonstrating the effectiveness of the proposed PG algorithm. 

In contrast, the fairness indexes\footnote{For illustration clarity, we do not depict the cluster throughput in Fig.\ref{fig:schemeA_reward1}b) and Fig.\ref{fig:schemeA_reward1}c). Nor in Fig.\ref{fig:schemeB_reward1}b) and Fig.\ref{fig:schemeB_reward1}c).}
shown in Fig.~\ref{fig:schemeA_reward1} that are obtained for the same network configurations when $\lambda = 1.0$ are 0.545, 0.559, and 0.560, respectively. 
Clearly, when the goal of the PG algorithm is to maximize system throughput, the throughput fairness is considerably penalized.

%These significantly lower fairness values reveal considerable inequity in transmission opportunities when the goal for access control is to maximize system throughput. 
However, when the PG algorithm intends to maximize cluster throughput fairness using $r^{(2)}$, 
the system throughput in the three studied network configurations when $\lambda = 1.0$ are, respectively, 8.26\%, 6.01\%, and 6.78\% lower 
than the ones obtained %when the PG algorithm is set to maximize the system throughput 
based on reward function $r^{(1)}$ under the same traffic load. %when $\lambda = 1.0$. 
In other words, cluster throughput fairness is achieved at the cost of merely a minor system throughput decrement.

\vspace{-3mm}  
%-----------------------------------------------------------------------------
\subsection{System and Cluster Throughput in Scheme B}
%-----------------------------------------------------------------------------
Having investigated the performance of the PG-driven access control mechanism
in terms of throughput and fairness in Scheme~A, 
we focus in this subsection on evaluating the impact that hashing has on system performance in Scheme B, with cluster $C_2$ in the CB mode and cluster $C_1$ in the SCF mode. 
As above, the metrics used to evaluate this impact are throughput and throughput fairness. 
The results are displayed in Fig.~\ref{fig:schemeB_reward1} and Fig.~\ref{fig:schemeB_reward2}, respectively.

%-----------------------------------------------------------------------------
\subsubsection{The benefit of hashing on system throughput} 
%-----------------------------------------------------------------------------
Let us first assess the benefit brought by the hash function 
when employing the radio resource allocation algorithm supported by the execution of a hash function at the devices, with %with d the periodic BS broadcast of hash seeds. 
the system throughput for all the three network configurations illustrated  
in Fig.~\ref{fig:schemeB_reward1}. 
In this figure, by PG(SA) (in red) and PG(SB) (in blue) we refer to the system throughput achieved by %\del{the proposed access control framework in} 
Scheme A and B, respectively, when reward function $r^{(1)}$ is deployed. 
As in Scheme A, the green curve in this figure represents a system throughput benchmark in Scheme B when $r^{(1)}$ is adopted. 
Here, the throughput \emph{upper-bound} benchmark has been obtained by exhaustive search of the access probabilities and the hash seeds, assuming that the BS has complete knowledge of the system state and the IDs of all ADTs  in the cell. 
%To obtain this throughput \textit{upper-bound} benchmark, we assume %leverage the fact 
%that the BS has complete knowledge of the system state and the IDs of all active devices in the cell. 
This knowledge allows the BS to configure the access probability and seed values in a way that maximize system throughput. %which is referred to as \textit{throughput benchmark}. 
However, it is noteworthy to clarify that this benchmark represents \textit{an ideal} upper-bound on system performance rather than an optimal solution, as it cannot be achieved in a real-life deployment scenario.
%of the proposed framework}. %However, note that such an access policy is not realistic in real-life scenarios.

%\del{It is computed assuming that the BS knows almost the full system state information, i.e., it has complete knowledge of the number of ADQ in each frame and the IDs of all devices in the system. This allows the BS to compute and broadcast access probabilities and hash-seed values that maximize system throughput.} {\color{red}Already explained.}

In Fig.~\ref{Hash_C0_Thp_8_8_4}), we still keep the curve labeled as \emph{wac} to show that the proposed PG-driven access control mechanism brings a huge benefit. 
Clearly, the system throughput in Scheme B is substantially higher then the one in Scheme A.  
In particular, it is 28.70\%, 26.65\%, and 24.30\% higher for the three network configurations when $\lambda = 1.0$, respectively, 
contributing significantly to improved total system throughput.

%------------------------------------------------------------------- 
 \begin{figure*}[ht]
  \centering
  \begin{subfigure}{0.32\textwidth}\vspace{1mm}
        \includegraphics[scale = 0.38]{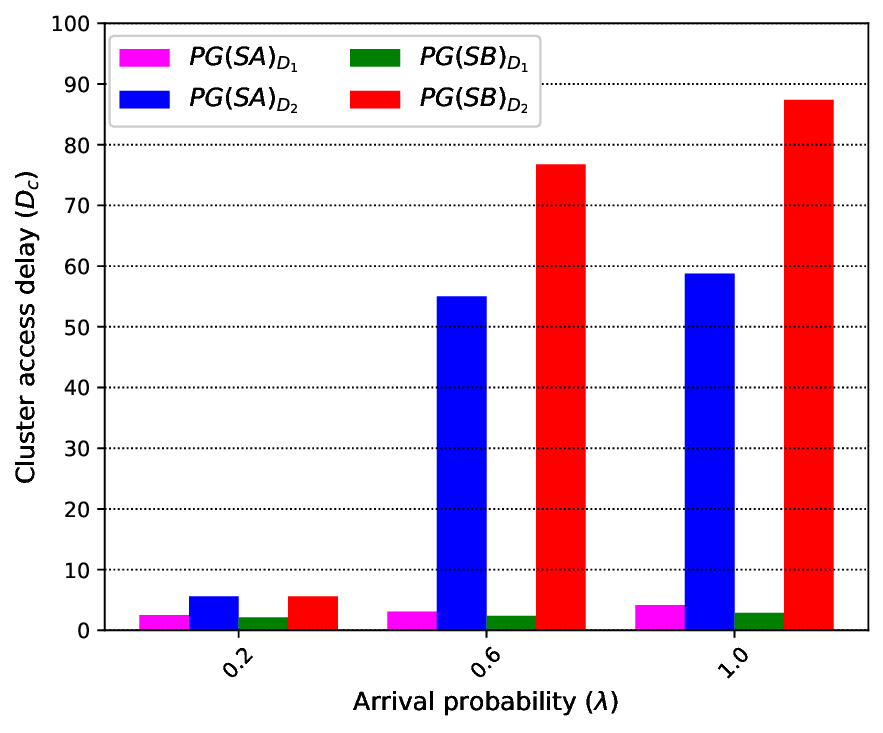} \vspace{-1mm}
        \caption{Delay $\{4;8+8\}$} \vspace{-1mm}
	\label{r1_Delay_8_8_4}
  \end{subfigure}
  \begin{subfigure}{0.32\textwidth}\vspace{1mm}
        \includegraphics[scale = 0.38]{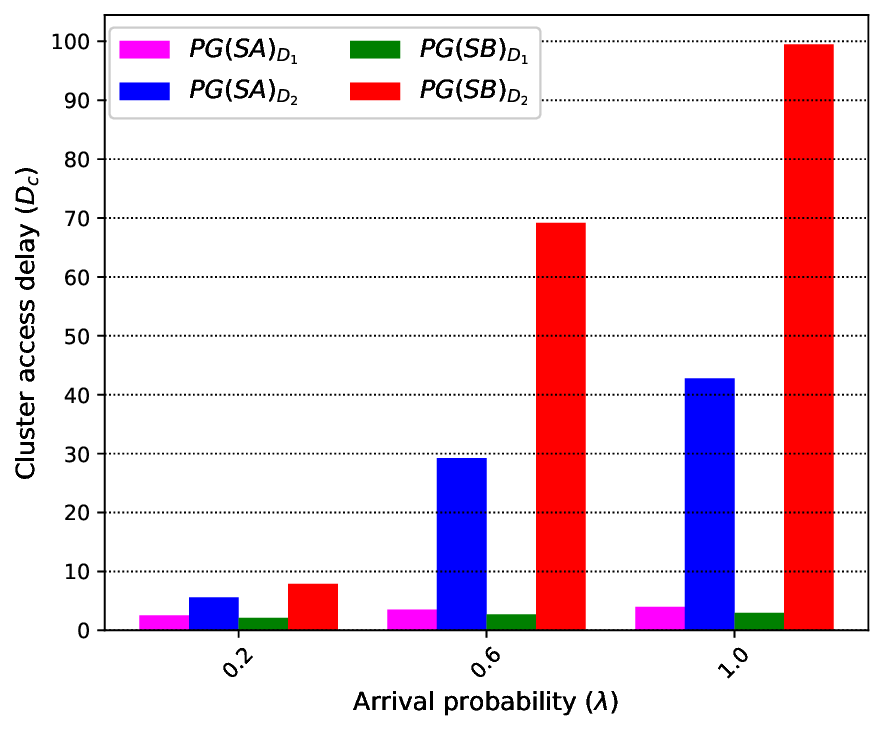} \vspace{-1mm}
        \caption{Delay $\{8;16+16\}$} \vspace{-1mm}
	\label{r1_Delay_16_16_8}       
  \end{subfigure} 
  \begin{subfigure}{0.32\textwidth}
        \vspace{-1mm}
        \includegraphics[scale = 0.38]{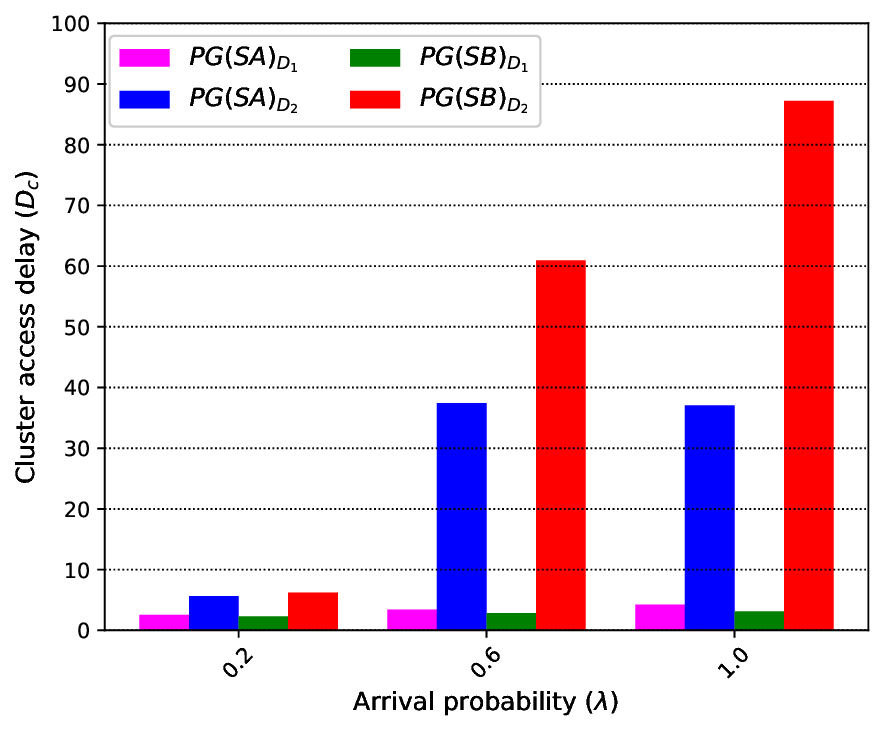} \vspace{-1mm}
        \caption{Delay $\{16;32+32\}$} \vspace{-1mm}
	\label{r1_Delay_32_32_16}     
  \end{subfigure} 
  \caption{Access delay when using reward function $r^{(1)}$ under light, medium, and heavy traffic load.}
% conditions, where $PG(SA)_{D1}$/$PG(SA)_{D2}$ and $PG(SB)_{D1}$/$PG(SB)_{D2}$ represent the achieved delay by scenario A and scenario B for $C_1$ and $C_2$ respectively. 
  \label{fig:schemeAB_r1_delays}
\vspace{-4mm}
\end{figure*}
%------------------------------------------------------------------- 
%------------------------------------------------------------------- 
 \begin{figure*}[ht]
  \centering
  \begin{subfigure}{0.32\textwidth}\vspace{1mm}
        \includegraphics[scale = 0.38]{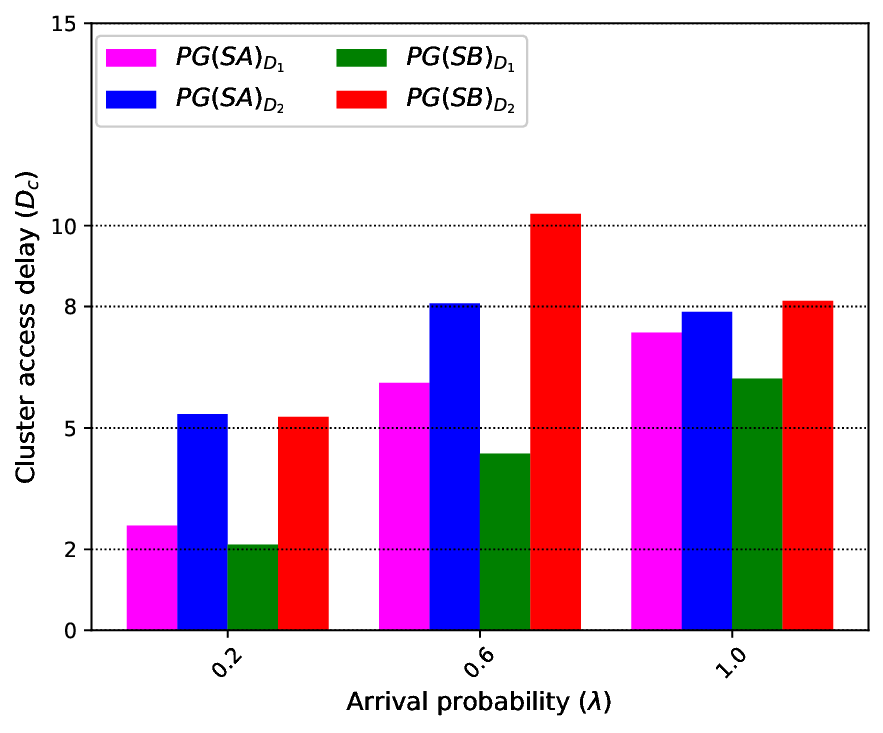} \vspace{-1mm}
        \caption{Delay $\{4;8+8\}$} \vspace{-1mm}
	\label{r2_Delay_8_8_4}
  \end{subfigure}
  \begin{subfigure}{0.32\textwidth}\vspace{1mm}
        \includegraphics[scale = 0.38]{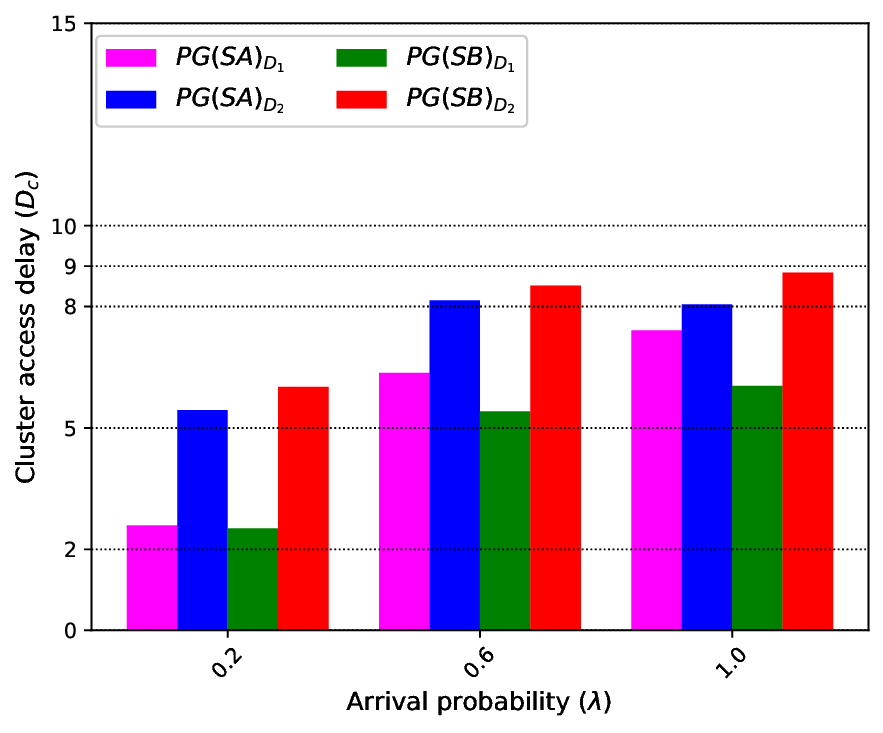} \vspace{-1mm}
        \caption{Delay $\{8;16+16\}$} \vspace{-1mm}
	\label{r2_Delay_16_16_8}       
  \end{subfigure} 
  \begin{subfigure}{0.32\textwidth}
        \vspace{-1mm}
        \includegraphics[scale = 0.38]{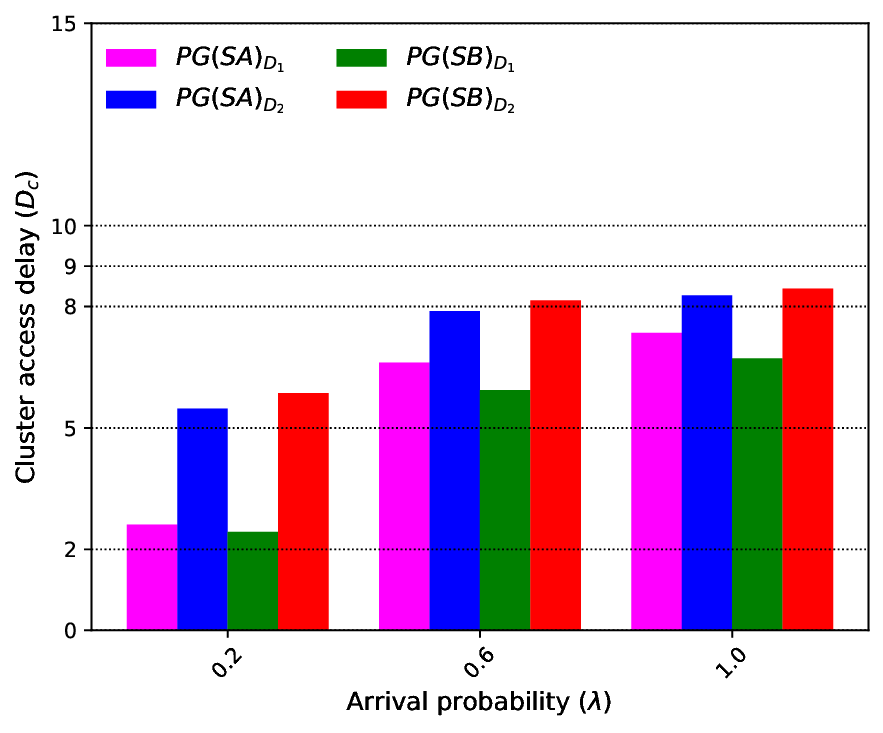} \vspace{-1mm}
        \caption{Delay $\{16;32+32\}$} \vspace{-1mm}
	\label{r2_Delay_32_32_16}     
  \end{subfigure} 
  \caption{Access delay when using reward function $r^{(2)}$ under light, medium, and heavy traffic load.}
	\label{fig:schemeAB_r2_delays}
% , where $PG(SA)_{D1}$/$PG(SA)_{D2}$ and $PG(SB)_{D1}$/$PG(SB)_{D2}$ represent the achieved delay by scenario A and scenario B for $C_1$ and $C_2$ respectively. 
\vspace{-6mm}
\end{figure*}
%------------------------------------------------------------------- 
%======================================================================
 
%-----------------------------------------------------------------------------
\subsubsection{Throughput fairness}
%-----------------------------------------------------------------------------
In Fig.~\ref{fig:schemeB_reward2}, we compare the cluster throughput fairness achieved in Scheme B 
when deploying reward functions $r^{(1)}$ and $r^{(2)}$. 
We denote by $PG^{r(n)}_{\gamma_m}$, $n=1,2$\,, $m= s,1,2$\,, the corresponding 
throughput achieved for the whole system, by cluster $C_1$ or $C_2$, respectively. %the proposed access control framework. 
 
Clearly, with $r^{(1)}$, higher system throughput and $C_1$ cluster throughput are achieved. 
On the other hand, when $r^{(2)}$ is adopted, we obtain much better throughput fairness. 
More specifically, the achieved fairness index values with $r^{(2)}$ are 0.98, 0.97, and 0.99 
for the three network configurations when $\lambda = 1.0$, respectively. 

However, the improved throughput fairness is obtained at the cost of lower system throughput. 
When deploying reward functions $r^{(2)}$, the observed system throughput reductions when $\lambda = 1.0$ are 20.90\%, 19.78\%, and 19.64\%\,, 
respectively, in comparison with the ones obtained when deploying $r^{(1)}$.
Note, however, that the system throughput with $r^{(2)}$ is higher in Scheme B than in Scheme A.
In general, which reward function to adopt depends on the service requirements. 

%{\color{red} We achieved significantly improved fairness values of 0.98, 0.97, and 0.99, albeit with a reduction in throughput when compared to Scheme B's reward function $r^{(1)}$. The throughput reductions were observed to be  20.90 \%, 19.78 \% , and  19.64 \% respectively. }

%{\color{blue} To achieve fairness while aiming to maximize system throughput for scheme B, we redefine reward function as in eq. \ref{reward_2}. Fig. \ref{fig:scheme2_reward2} outline the variation in system/cluster throughput with $\lambda \in [0.1,1]$ with system configuration \( \{4;8+8\} \), \( \{8;16+16\} \), and \( \{16;32+32\} \) }

%When fairness is considered, C2 devices achieve short delay.  
%====================================
%Fig.\ref{fig:scheme2_delays_throughput} and Fig.\ref{fig:scheme2_delays_fairness} present the variation in average access delay under low, medium, and high traffic loads, considering different network configurations. These variations are analyzed with different RL algorithms in the context of the reward functions defined in (\ref{reward 1}) and (\ref{reward_2}) and $wac$, offering a comprehensive comparison of the impact of traffic load and network configuration on average access delay per cluster. 

%==============================================================

%-----------------------------------------------------------------------------------
\vspace{-4mm}
\subsection{Access Delay} %: The Effect of Reward Function and Access Scenario
%==============================================================
%------------------------------------------------------------------- 
\subsubsection{Access delay when reward function $r^{(1)}$ is adopted}
%------------------------------------------------------------------- 
As defined in~(\ref{reward 1}), the main goal of reward function $r^{(1)}$ is to maximize system throughput, 
regardless of the achieved throughput fairness index. 
In pursuit of this goal, the PG algorithm assigns a higher access probability to $C_1$ devices than to $C_2$ devices, 
as the first ones are closer to the BS than the second ones.   
Then, $C_1$ devices achieve lower and more stable delays under all traffic load conditions and across all network configurations,  
as shown in the pink and green bars in all three sub-figures in Fig.~\ref{fig:schemeAB_r1_delays}. 
On the other hand, devices in $C_2$, with a lower access probability, 
perceive longer delays per successfully SDP transmission. %in successfully transmitting their packets. 

When comparing the access delay achieved in Schemes A and B, 
it is clear that $C_2$ devices achieve comparatively longer delays in Scheme B than in Scheme A. 
As explained above, the RL algorithm assigns an even higher access probability to $C_1$ devices in Scheme B than in Scheme A, as $C_1$ devices in Scheme B adopt the SCF access mode, resulting in substantially improved successful detection rate. This leads to a small delay reduction for packets from $C_1$ devices in Scheme B than in Scheme A. 
However, packets from $C_2$ devices in Scheme B experience a substantially longer delay than in Scheme A.
%This is because in Scheme B, $C_1$ devices deploy the SCF mode, that is boosted by the operation of the hashing function.  Then, $C_1$ devices perceive less intra-cluster interference, leading to a higher SDP detection rate at the BS. This result is caused by the fact that the PG algorithm deploys $r^{(1)}$, leading to an additional access probability increment being assigned to $C_1$ devices in detriment to an even lower access probability being assigned to $C_2$ devices. As a consequence, $C_1$ devices operating in Scheme B perceive an even shorter access delay than in Scheme A, at the cost of a longer access delay for $C_2$ devices. 

%----------------------------------------------------------------------- 
\subsubsection{Access delay when reward function $r^{(2)}$ is adopted}
%----------------------------------------------------------------------- 
Different from $r^{(1)}$, reward function $r^{(2)}$ defined in (\ref{reward_2}) aims to achieve throughput fairness across clusters. 
%It seeks more equally distributed throughput among clusters, thereby, 
Accordingly, balanced access delays among devices from different clusters may be achieved.
%caused by unequal opportunities and leading to comparatively small delay differences for $C_1$ and $C_2$ devices, no matter which access scheme is employed. 
%evenly contribution of access delays among clusters  

When comparing the results in Fig.~\ref{fig:schemeAB_r1_delays} and Fig.~\ref{fig:schemeAB_r2_delays}, 
we reveal that much shorter access delay differences between clusters have been achieved by using reward function $r^{(2)}$ 
than when using $r^{(1)}$, in both evaluation schemes and for all the investigated network configurations and traffic load conditions. 
Take network configuration $ \{8;16+16\} $ and medium traffic load, $\lambda=0.6$\,, as an example. 
In Scheme A, $C_1$ and $C_2$ devices achieve an access delay  
(marked in pink and blue in Fig.~\ref{r1_Delay_16_16_8}) and Fig.~\ref{r2_Delay_16_16_8}), respectively) of 3.51 and 29.22 frames with $r^{(1)}$, 
versus 6.36 and 8.15 frames with $r^{(2)}$.  
However, this delay fairness improvement is achieved at the expense of $C_1$ devices experiencing longer delay. % without much performance degradation. 
In other words, 3.58 times shorter delay has been experienced by $C_2$ devices 
at the cost of 1.81 times longer delay experienced by $C_1$ devices. 
With the same network configuration $\{8;16+16\}$ and traffic load $\lambda=0.6$ but using Scheme B, 
where $C_1$ and $C_2$ devices deploy the SCF and CB modes, respectively, 
devices in $C_1$ enjoy 2.00 times shorter access delay than $C_2$ devices at the cost of 8.12 times longer access delay for $C_2$ devices  
(marked in green and red in Fig.~\ref{r1_Delay_16_16_8}) and Fig.~\ref{r2_Delay_16_16_8}), respectively). 

\vspace{-3mm}
%==============================================================
\subsection{Device Energy Consumption}
%==============================================================
In Table~\ref{tab:energy_scheme:A,reward 1,2}, we present the average energy consumed by a device per successfully SDP transmission, 
where $E_1$ and $E_2$ represent the per device energy consumption averaged across all the devices in $C_1$ and $C_2$ and frames throughout the simulation, respectively. 
In general, a $C_1$ device consumes less energy than a $C_2$ device does for all the investigated network configurations and traffic load conditions. 
Clearly, $C_1$ devices, being closer to the BS, require less number of attempts to complete a successful SDP transmission than $C_2$ devices do.  

For a fixed traffic load, the energy consumed by both $C_1$ and $C_2$ devices increases moderately with network size. 
Although the ratio  of the number of radio resources (time slots) in a frame per device is maintained identically for all the studied network configurations, 
more active devices per frame in a larger network induce higher interference. 
%Clearly, this is due to the fact that in larger networks  

% ----------------------------------------------------------------------------------------
\begin{table*}[t]
\caption{Per device energy consumption (in mJ) in Scheme A} %\vspace{-1mm} %under light, medium, and heavy traffic load conditions: Reward function $r^{(1)}$ vs. $r^{(2)}$} 
\centering
\label{tab:energy_scheme:A,reward 1,2}
\centering
\renewcommand{\arraystretch}{1} % Increase row height by 1.5 times
\begin{tabular}{|c|*{12}{>{\centering\arraybackslash}m{1cm}|}} % Use fixed-width columns with centering
\hline
\multirow{3}{*}{$\lambda$} & \multicolumn{6}{c|}{Reward function $r^{(1)}$} & \multicolumn{6}{c|}{Reward function $r^{(2)}$} \\ \cline{2-13}
& \multicolumn{2}{c|}{$\{4; 8+8\}$} & \multicolumn{2}{c|}{$\{8;16+16\}$} & \multicolumn{2}{c|}{$\{16;32+32\}$}
& \multicolumn{2}{c|}{$\{4; 8+8\}$} & \multicolumn{2}{c|}{$\{8;16+16\}$} & \multicolumn{2}{c|}{$\{16;32+32\}$} \\ \cline{2-13}
& $E_1$ & $E_2$ & $E_1$ & $E_2$ & $E_1$ & $E_2$ & $E_1$ & $E_2$ & $E_1$ & $E_2$ & $E_1$ & $E_2$ \\ \hline
$0.2$ & $8.75$ & $18.03$ & $9.30$ & $18.71$ & $10.21$ & $19.44$ & $8.80$ & $18.12$ & $9.34$ & $18.74$ & $10.26$ & $19.51$  \\ \hline
$0.6$ & $10.87$ & $23.91$ & $12.20$ & $23.37$ & $13.84$ & $26.01$ & $11.29$ & $24.30$ & $12.53$ & $25.77$ & $12.98$ & $26.03$ \\ \hline
$1.0$ & $12.53$ & $25.00$ & $13.66$ & $27.34$ & $15.41$ & $28.03$ & $11.89$ & $25.90$ & $12.28 $& $26.31$ & $13.19$ & $27.19$ \\ \hline
\end{tabular}
\end{table*}
\begin{table*}[t]
\caption{Per device energy consumption (in mJ) in Scheme B} %\vspace{-1mm} %under light, medium, and heavy traffic load conditions: Reward function $r^{(1)}$ vs. $r^{(2)}$  } 
\centering
\label{tab:energy_scheme:B,reward 1,2}
\centering
\renewcommand{\arraystretch}{1} % Increase row height by 1.5 times
\begin{tabular}{|c|*{12}{>{\centering\arraybackslash}m{1cm}|}} % Use fixed-width columns with centering
\hline
\multirow{3}{*}{$\lambda$} & \multicolumn{6}{c|}{Reward function $r^{(1)}$} & \multicolumn{6}{c|}{Reward function $r^{(2)}$} \\ \cline{2-13}
& \multicolumn{2}{c|}{$\{4; 8+8\}$} & \multicolumn{2}{c|}{$\{8;16+16\}$} & \multicolumn{2}{c|}{$\{16;32+32\}$}
& \multicolumn{2}{c|}{$\{4; 8+8\}$} & \multicolumn{2}{c|}{$\{8;16+16\}$} & \multicolumn{2}{c|}{$\{16;32+32\}$} \\ \cline{2-13}
& $E_1$ & $E_2$ & $E_1$ & $E_2$ & $E_1$ & $E_2$ & $E_1$ & $E_2$ & $E_1$ & $E_2$ & $E_1$ & $E_2$ \\ \hline
0.2 & 7.28 & 17.29 & 8.15 & 19.55 & 9.57 & 19.96 & 7.54 & 17.76 & 9.13 & 20.58 & 11.84 & 26.70 \\ \hline
0.6 &  8.90 & 32.70 & 10.77 & 29.01 & 12.75 & 27.89 & 9.30 & 25.47 & 10.30 & 27.35 & 11.78 & 27.19 \\ \hline
1.0 & 9.30 & 33.10 & 11.43 & 39.25 & 13.97 & 36.52 & 9.24 & 24.80 & 10.94 & 27.66 & 12.09 & 27.86 \\ \hline
\end{tabular} \vspace{-3mm}
\end{table*}
% -----

%----------------------------------------------------------------------- 
\subsubsection{Energy consumption, Scheme A versus Scheme B}
%----------------------------------------------------------------------- 
Comparing the energy consumption in Schemes A and B for a given network size and load, 
we reveal that, in Scheme B, $C_1$ device consumption is lower, while $C_2$ device consumption is higher than in Scheme A.  
In Scheme B, $C_1$ devices employ the SCF mode, enjoy less intra-cluster interference and, 
then, less transmission attempts per successful SDP transmission, resulting in less energy consumption. 
On the other hand, the fact that $C_1$ devices are closer to the BS creates higher interference on $C_2$ devices, than vice versa. %$C_2$ devices on $C_1$ ones. 
This effect is more acute when the BS runs $r^{(1)}$ as, in Scheme B, the PG algorithm perceives a higher 
successful detection rate for $C_1$ devices and increases its access probability with respect to Scheme A. 

Take network configuration $\{8;16+16\}$, running $r^{(1)}$ and medium traffic load, $\lambda=0.6$, as an example.  
Let $N_{tot}^X(r^{(i)},C_i)$, $X=A,B$\,, $i=1,2$\,, be the average number of required transmission attempts per successful SDP transmission 
for a $C_i$ device in Scheme $X$ when the BS runs $r^{(i)}$. 
We obtain that $N_{tot}^A(r^{(1)},C_1)=3.17$ and $N_{tot}^B(r^{(1)},C_1)=2.62$. 
Then, the energy consumed by $C_1$ devices in Scheme B is $11.7\%$ lower than in Scheme A. 
We show also that $N_{tot}^A(r^{(1)},C_2)=6.79$ and $N_{tot}^B(r^{(1)},C_2)=8.46$. 
Then, the energy consumed by $C_2$ devices in Scheme B is $24.6\%$ higher than in Scheme A. 

As mentioned previously, the energy consumed by $C_2$ devices is higher than the one consumed by $C_1$ devices.  
For $\{8;16+16\}$ and $\lambda=0.6$, 
we find that, in Scheme A, $N_{tot}^A(r^{(1)},C_1)=3.17$ and $N_{tot}^A(r^{(1)},C_2)=6.79$.   
Then, the energy consumed by $C_2$ devices is $91.6\%$ higher than the one consumed by $C_1$ devices.   
We also illustrate that $N_{tot}^B(r^{(1)},C_1)=2.62$ and $N_{tot}^B(r^{(1)},C_2)= 8.46$, and the energy consumed by $C_2$ devices is $169.4\%$ higher than the one consumed by $C_1$ devices in Scheme B. 

%----------------------------------------------------------------------- 
\subsubsection{Energy consumption, $r^{(1)}$ versus $r^{(2)}$}
%----------------------------------------------------------------------- 
In Scheme A, the energy consumed by $C_1$ and $C_2$ devices does not change significantly no matter the BS runs $r^{(1)}$ or $r^{(2)}$, 
except for large network sizes, where the energy consumption slightly decreases when the BS runs $r^{(2)}$ instead of $r^{(1)}$. 
In Scheme B, the energy consumed by $C_1$ devices follows a similar trend. 

The energy consumed by $C_2$ devices in Scheme A increases when the BS runs $r^{(2)}$ instead of $r^{(1)}$.  
However, in Scheme B, it decreases when the BS runs $r^{(2)}$ instead of $r^{(1)}$.  
When the BS runs $r^{(2)}$, the objective of the PG algorithm is to maximize cluster throughput fairness, 
and this is achieved by reducing the access probability to $C_1$ devices. 
In Scheme A, the higher power level of $C_1$ with respect $C_2$ SDP signals received by the BS,  
together with the random time slot selection of active $C_1$ and $C_2$ devices, 
have a higher impact on the inter-cluster interference perceived by $C_2$ devices than 
the fact that $C_1$ devices are assigned a lower access probability. 
However, in Scheme B, $C_1$ devices deploy the SCF access mode that helps to evenly spread SDP transmissions over different time slots. 
As a consequence, $C_2$ devices perceive less inter-cluster interference and their successful detection rate increases 
when the BS runs $r^{(2)}$ instead of $r^{(1)}$. 
 
Take network configuration $\{8;16+16\}$ and medium traffic load, $\lambda=0.6$, as an example.  
We find that, in Scheme A, $N_{tot}^A(r^{(1)},C_1)=3.34$ and $N_{tot}^A(r^{(2)},C_1)=3.43$. 
Then, the energy consumption of $C_1$ devices when the BS runs $r^{(2)}$ is $2.7\%$ lower than when running $r^{(1)}$. 
We reveal further that, in Scheme B, $N_{tot}^B(r^{(1)},C_1)=2.62$ and $N_{tot}^B(r^{(2)},C_1)=2.51$. 
Then, the energy consumption of $C_1$ devices when the BS runs $r^{(2)}$ is $4.4\%$ lower than when running $r^{(1)}$. 

Moreover, we illustrate that, in Scheme A, $N_{tot}^A(r^{(1)},C_2)=6.79$ and $N_{tot}^A(r^{(2)},C_2)=7.36$. 	
Then, the energy consumption of $C_2$ devices when the BS runs $r^{(2)}$ is $10.3\%$ higher than when running $r^{(1)}$. 
For comparison, we obtain in Scheme B $N_{tot}^B(r^{(1)},C_2)=8.46$ and $N_{tot}^B(r^{(2)},C_2)=7.56$.
Then, the energy consumption of $C_2$ devices when the BS runs $r^{(2)}$ is $5.7\%$ lower than when running $r^{(1)}$. 

In summary, the most favorable configuration for reducing energy consumption is Scheme B with the BS running $r^{(2)}$. 
Here, the energy consumed by $C_1$ devices is the lowest one.
While the energy consumed by $C_2$ devices is approximately the same as in Scheme A, either running $r^{(1)}$ or $r^{(2)}$, it is
lower than the one consumed in Scheme B with the BS running $r^{(1)}$.  

%------------------------------------------------------------------- 
 \begin{figure*}[ht]
  \centering
  \begin{subfigure}{0.32\textwidth}\vspace{1mm}
        \includegraphics[scale = 0.41]{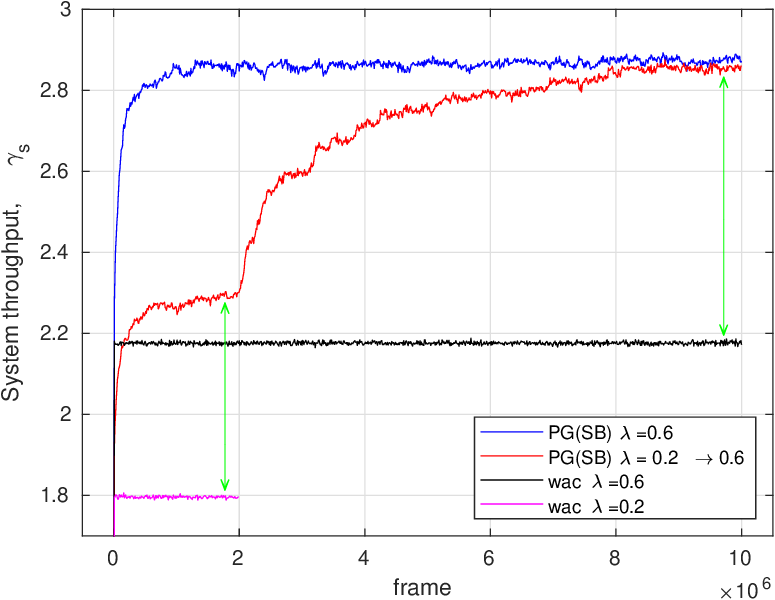} \vspace{-1mm}
        \caption{System throughput over time from cold start and abrupt transition: Reward function $r^{(1)}$} \vspace{-1mm}
	\label{fig:convergence1}       
  \end{subfigure} 
  \begin{subfigure}{0.32\textwidth}
        \vspace{-1mm}
        \includegraphics[scale = 0.41]{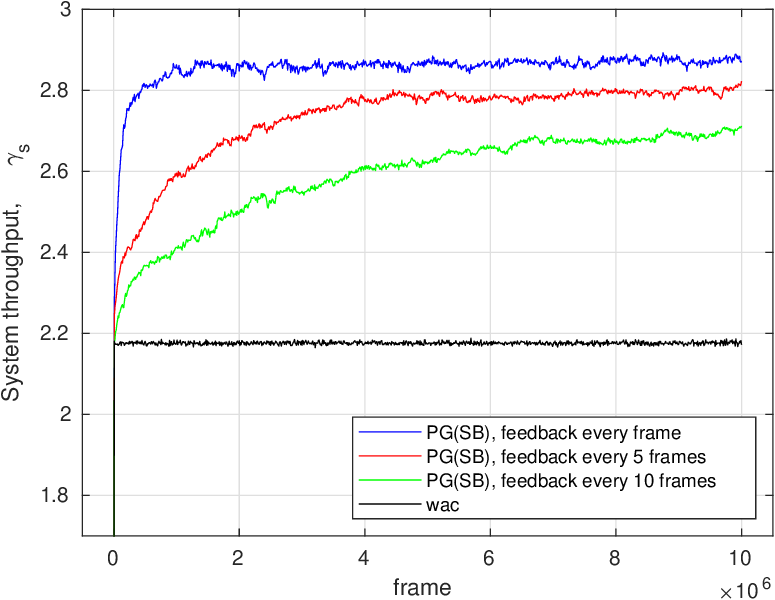} \vspace{-1mm}
        \caption{System throughput over time with 3 different updating intervals: Reward function $r^{(1)}$} \vspace{-1mm}
	\label{fig:convergence2}     
  \end{subfigure} 
   \begin{subfigure}{0.32\textwidth}\vspace{1mm}
        \includegraphics[scale = 0.38]{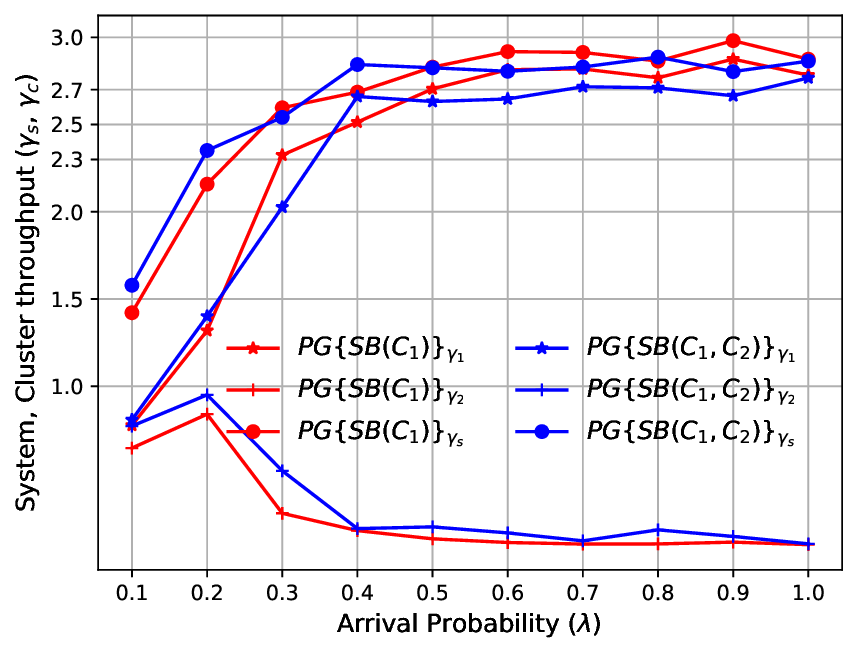} \vspace{-1mm}
        \caption{System and cluster throughput: Hashing applies to both clusters: Reward function $r^{(2)}$} \vspace{-1mm}
	\label{hash_coc1}
  \end{subfigure}
  \caption{Scheme B for network configuration $\{4;8+8\}$: (a) System throughput from a cold start; (b) System throughput with different updating intervals; (c) Throughput: Hashing for both.} 
% conditions, where $PG(SA)_{D1}$/$PG(SA)_{D2}$ and $PG(SB)_{D1}$/$PG(SB)_{D2}$ represent the achieved delay by scenario A and scenario B for $C_1$ and $C_2$ respectively. 
  \label{fig:cold_interval_hashing}
\vspace{-6mm}
\end{figure*}
%-------------------------------------------------------------------

\vspace{-3mm}
\section{Feasibility and Operability} %of the Framework}
%\section{FEASIBILITY OF THE FRAMEWORK} %\vspace{-2mm}
\label{Sec:feasibility_discussions}
%===============================================================
In this section, we further explore six other aspects that may have an effect on or are related to the feasibility and operability of the developed RL4SCF framework.

\vspace{-3mm} 
%----------------------------------------------------------------------- 
\subsection{Optimality}
%\label{subsec:transmission_principle}
%----------------------------------------------------------------------- 

By comparing the system throughput achieved in Scheme A with the system throughput benchmark in Fig.~\ref{fig:schemeA_reward1}, % \del{and Fig.~\ref{fig:schemeA_reward2}}, 
we conclude that quasi-optimal performance has been achieved by the proposed PG algorithm. %using this scheme.
When observing the corresponding results for Scheme B in Fig.~\ref{fig:schemeB_reward1}, % \del{and Fig.~\ref{fig:schemeB_reward2}}, 
%on the other hand, we notice that sub-optimal performance is achieved.
we notice that the achieved system throughput is below
the benchmark for some network configurations. %As the access policy to achieve this benchmark is not realizable, further work is needed to identify new performance enhancing features, if they exist.
This result is in accordance with the operation principle of Scheme B, as multi-dimensional 
policies have to be learned online simultaneously when running both SCF and CB, %in this case, 
making the optimization problem more challenging with a larger network size. 

Considering that the PG-agent in our framework is purely %data-driven 
model-free and actions are taken 
solely based on the agent's partial knowledge on system state, 
we ascertain that feasibility, operability, and excellent performance have been achieved by the proposed PG-driven access control and resource allocation mechanism for SDP transmissions. 

\vspace{-3mm}
%----------------------------------------------------------------------- 
\subsection{Convergence} 
%\label{subsec:}
%----------------------------------------------------------------------- 
To provide an insight into the convergence time of the RL-enabled control mechanism, we reveal in Fig.~\ref{fig:convergence1}) the evolution of the average system throughput over time, obtained based on Scheme B, from a cold start and also from an abrupt transition of the traffic load.
The average throughput is measured in a time window of $10^4$ frames, and the plot illustrates the average values based on $10$ runs.
The blue line is the throughput from a cold start, with the policy parameters $\vec{\theta}_i$ and $\vec{\phi}_i$ set to $0$, and with the arrival probability to all devices set, for consistency and uniformity, to $\lambda = 0.6$.
The red line is the system throughput from a cold start with %the arrival probabilities to all devices set to 
$\lambda = 0.2$ until frame $2\cdot 10^6$, the moment at which 
%all arrival probabilities are set to 
$\lambda = 0.6$. 

From a cold start, more than $90\%$ of the gain is achieved around the $2\cdot10^5$  frame, and after that the gain converges slowly to the maximum.
This can be seen in the blue line, for the high load case, and in the red line up to frame $2\cdot 10^6$, for low load.
The green arrows show the maximum throughput gains relative to transmission without access control.
As can be observed, the system takes longer time to converge from a previously learned behavior than from a start with the policy parameters set to zero.
However, it is worth noting that this experiment examines an extreme case, in which the load changes instantaneously on all devices from a very low value to a very high value. In a more realistic scenario, the system load would likely change more smoothly. 

\vspace{-3mm}
%----------------------------------------------------------------------- 
\subsection{Periodicity of BS Broadcasts}
%\label{subsec:transmission_principle}
%----------------------------------------------------------------------- 
%In the above descriptions about RL4SCF, 
In the numerical results presented above in Section \ref{sec:results}, 
%we have presented the operation of the RL-based access control mechanism exemplified based on the per-frame observation of the BS, {\color{blue}i.e., %access probabilities and hash seeds are updated every frame.}
the BS broadcasts an access probability and a hash seed per cluster 
once every $\mathcal{K}= 1 $ frame. %in every frame. 
However, it is worth clarifying that the BS broadcast periodicity in the RL4SCF framework can be set to any other value by appropriately configuring $\mathcal{K}$.
%in a flexible manner and this interval can be set to other positive integer values.}
%is a configurable parameter that
%Hereby, it is worth clarifying that the duration of a cycle to perform RL-enabled dynamic access control in our framework can be configured in a flexible manner.

%In all the results shown so far, we have assumed that access control is performed on a frame-by-frame basis, that is, access probabilities and  hash seeds are updated every frame. 
As an example, we investigate in Fig.~\ref{fig:convergence2}) how a longer updating and broadcast interval would affect performance and convergence time, by illustrating the evolution of the average system throughput over time in Scheme B, for updating intervals configured as $ \mathcal{K} \in \{1,5,10\}$ frames respectively. The results reveal that the algorithm converges faster with a shorter updating and broadcast interval, i.e., a smaller $ \mathcal{K}$.

%e.g., at the periodicity of the master information block (MIB) or the system information block (SIB) cycle (80, 160 ms, or another value)~\cite{TS38331}.{\color{red}This paragraph is added based on a comment of Jorge last week.\\ I do not think it is a good idea to tie or framework to this or any other numbers.\\  \emph{The way The device gets to know the radio resources for the RA procedure from system information messages, in a similar way to non RA-SDT devices. However, the RA radio resources for SDT and non SDT devices are kept separate; that is, these device types do not interfere with each other in random access.}} 

\iffalse
\begin{figure}
    \centering \vspace{-1mm}
    \includegraphics[scale = 0.48]{Figures/convergence1.eps} \vspace{-2mm}
    \caption{System throughput over time from a cold start and from an abrupt transition of the traffic load, in Scheme B, network size $\{4;8+8\}$ and reward function $r^{(1)}$.     }\vspace{-3mm}
    \label{fig:convergence1}
\end{figure}

\begin{figure}
    \centering \vspace{-1mm}
    \includegraphics[scale = 0.48]{Figures/convergence2.eps} \vspace{-2mm}
    \caption{System throughput over time from a cold start for several updating intervals, in Scheme B, network size $\{4;8+8\}$ and reward function $r^{(1)}$.     }\vspace{-3mm}
    \label{fig:convergence2}
\end{figure}
\fi

\vspace{-4mm}
%----------------------------------------------------------------------- 
\subsection{Complexity and Scalability} 
%----------------------------------------------------------------------- 
In a general sense, the complexity of the PG algorithm (Alg.~\ref{alg:PGradient}) is affected by two factors related to network size, as explained below:  
\begin{itemize}
\item[-] Number of time slots. Let $\mathcal{S}$ be the state space and $\left|\mathcal{S}\right|$ its cardinality. 
Then, $\left|\mathcal{S}\right|= fL+1$, with $L$ being the number of time slots allocated for SDP transmissions 
and $f$ being the maximum number of SDP successful detections per time slot. 
Note that $f$ will take a small value as in the studied network configurations. %\del{where we set $f=2$.} {\color{red}JR: no need to say how much is f} 
%Should be 3, if you agree with our comment on page 5, IV-B.}  
For each cluster $C_i$, the access probability policy $\pi_i(\cdot|s,\vec{\theta}_i)$ 
has to maintain and update a parameter vector $\vec{\theta}_i$ with $\left|\mathcal{S}\right|$ elements.
Each seed policy $\tau_i(\cdot|s,\vec{\phi}_i)$ 
has to maintain and update a parameter vector $\vec{\phi}_i$ with $\left|\mathcal{S}\right|\times q$ elements, with 
$q$ being the number of candidate hash seeds. 
Also, the parameter vector $\vec{\omega}$ in the state-value function $V(s,\vec{\omega})$ has $\left|\mathcal{S}\right|$ elements.
\item[-] Number of clusters $C$. For each cluster, two policies are required:  $\pi_i$ for the access probability and $\tau_i$ for the seed for cluster $C_i$.
\end{itemize}

Additionally, each of these policies is composed of a set of \emph{independent} policies, one for each cluster. Thus, the domain of each of the functions to be learned (the policies) is discrete and one-dimensional, with size $n$ (number of states), which in turn depends solely on the number of active devices. 
Therefore, the complexity of Alg.~\ref{alg:PGradient}, in terms of memory size and processing time, \emph{grows only linearly} both with the number of time slots 
and with the number of clusters, demonstrating the scalability of the developed RL-enabled mechanism for access control and data transmission of uplink IoT traffic.

In contrast, \emph{conventional RL methods} require to learn and use a state-action-value function, and the domain dimension of this function grows with the number of clusters. Furthermore, the action space must be discretized. Therefore, the number of elements in the domain of the state-value function would be $n \times P^C \times q^C$, where %$C$ is the number of clusters, 
$q$ is the number of candidate seeds and $P$ is the number of access probabilities resulted in from the discretization of the interval $[0\; 1]$. Furthermore, the selection of actions requires a search, which is generally not efficient, in the state-action-value function. Consequently, these methods exhibit higher complexity in terms of both memory size and processing time, and this complexity potentially grows exponentially with the size of the problem, limiting their scalability. 

\vspace{-3mm}
%----------------------------------------------------------------------- 
\subsection{Applying Hashing to C2 Devices}
%\label{subsec:transmission_principle}
%-----------------------------------------------------------------------
We are also interested in investigating the performance of the proposed framework when the SCF mode is applied to $C_2$ devices as well. 
This is because we expect that a hashing-based slot selection by $C_2$ devices will lead to significantly lower intra-cluster interference %among $C_2$ transmissions 
and slightly less inter-cluster interference to $C_1$ transmissions.  

When reward function $r^{(1)}$ is adopted, the results are almost identical to those shown in Fig.~\ref{fig:schemeB_reward1}
that are obtained when only $C_1$ devices deploy the SCF mode. 
The reason for this behavior is straightforward. 
The BS receives weaker signals from $C_2$ devices than from $C_1$ devices. 
Accordingly, %SDP from $C_2$ devices are received as weak signals at the BS, 
the successful SIC detection rate of SDPs from $C_2$ devices is largely dependent on the absence of any concurrent transmission from $C_1$ devices. 
Then, even when the activities of $C_2$ devices produce lower intra- and inter-cluster interference, 
its impact is faded away by the signal strength of $C_1$ devices. 
Clearly, when $r^{(1)}$ is adopted, $C_1$ devices will still be assigned a much higher access probability 
than $C_2$ devices.  

In this line, we expect that when reward function $r^{(2)}$ is adopted for the purpose of improving throughput fairness, 
the impact of configuring $C_2$ devices with the SCF mode will be more noticeable. 
Taking network size $\{4;8+8\}$ as an example, we illustrate in Fig.~\ref{hash_coc1}) the obtained cluster and system throughput.
The curves in blue (labeled as $PG\{SB(C_1,C_2)\}\gamma$, etc.) represent the throughput when both $C_1$ and $C_2$ devices run the SCF mode, whereas the curves in red (labeled as $PG\{SB(C_1)\}\gamma$, etc.) represent the throughput when only $C_1$ devices run SCF.   
With $r^{(2)}$, $C_1$ devices are assigned lower access probabilities than with $r^{(1)}$. 
Then, with lower concurrent transmission from any $C_1$ device, slot allocation through hashing substantially %apparently 
reduces intra-cluster interference and 
increases the successful SDP detection rate from $C_2$ devices, as shown in the $\gamma_2$ curves in Fig.~\ref{hash_coc1}). 

Although not shown explicitly herein, %for space limitation, 
an additional benefit of running the SCF mode in both clusters is 
a reduction of energy consumption. On average, $C_1$ and $C_2$ devices observe a relative energy consumption reduction of 
$20.5\%$ and $18.4\%$, respectively. 

\vspace{-3mm}
\subsection{Applicability to Multiple Clusters} 
This study focuses on a network scenario considering that devices in a cell covered by the same beam are confined into two clusters and our access control and SDP transmission schemes deal with  both intra- and inter-cluster interference. When more than two clusters exist in the same cell, inter-cluster interference would increase significantly. Specifically, the performance of devices located in the farthest cluster(s) from the BS would likely deteriorate due to higher path loss combined with increased interference. As a consequence, it would become much more difficult for the SIC procedure to effectively distinguish and detect data packets successively. 

%Investigating scenarios involving more than two clusters would provide additional insights on the applicability of our framework.  
Such a problem in NOMA networks is indeed well known in the research community. To solve this problem, various \emph{user pairing} and \emph{clustering} strategies have been proposed (see e.g., \cite{rajasekaran2023neural}\cite{ren2019em}\cite{TWC2016}\cite{IEEA2019}\cite{cmm} 
\cite{Shahini2019}). Once coupled with a proper cluster paring algorithm, the proposed framework is inherently scalable and can be extended to accommodate multiple clusters in larger network deployments. However, developing any other cluster pairing strategy or algorithm beyond the location-based clustering is outside the scope of this paper.

%================================================================================================================
\vspace{-3mm}
\section{Conclusions and Future Work}
\label{Sec:Conclusions}
%================================================================================================================
In this paper, we have presented an RL-enabled framework for SDP transmission tailored to uplink traffic in clustered NOMA-facilitated IoT networks, 
supporting hybrid contention-based and semi-contention-free access modes. 
As the core component of the framework, the BS, as an RL-agent, performs online learning through a policy gradient algorithm.
It computes access probabilities for both access modes to maximize system throughput or achieve throughput fairness among clusters. 
Also, it computes hash seeds to support the semi-contention-free access mode operation. 
When a device operates in the semi-contention-free mode, it reduces both intra- and inter-cluster interference. 
To operate the RL4SCF framework, no assumption on the state of the system (number of active devices) is required at the BS. 
Nor is it necessary for the BS and devices to %acquire channel state information or 
perform any protocol handshake prior to a data transmission.  
IoT devices only need to execute a lightweight hash function  
and to perform simple computations based on the seed and access probability that they receive from the BS periodically. 

By illustrating various numerical results in terms of four performance metrics, % obtained through extensive simulations, 
we showcase not only the operability and efficiency but also the scalability and feasibility of the RL-enabled solution 
for random access and SDP transmission in B5G IoT networks. 
The findings of this study include: 
1) It is beneficial to integrate SCF with CB for SDP transmission and which combination to apply depends on service requirements; 
%2) Data-driven model-free RL-based approaches excel model-based methods for resource allocation and data transmission in complex environments; 
2) The proposed framework exhibits robust and stable performance, being able to achieve quasi-optimal or excellent
%very good 
performance in the investigated network configurations; and
3) The complexity of our developed framework is low and the access modes scale well with network size. For devices, light computation capacity suffices.  
As our future work, we regard integrating more precise physical layer models, developing cluster pairing algorithms, and applying other RL algorithms into the framework as potential directions.

%\appendices{}
\vspace{-3mm}
\appendix[Summary of Notations and Descriptions]{}\centering

%\section{List of Notations}\label{notations}

\begin{table}[t]\small
    %\caption{{\color{blue}Summary of Notations and Descriptions}}
    \centering
% {\color{blue}
    \begin{tabular}{|p{13mm}|p{67mm}|} % Set column widths here
        \hline
        \textbf{Notation} & \textbf{Description} \\ \hline
        $C$ & Total number of cluster \\ \hline
        $C_i$ & Cluster $i$ \\ \hline
        $D^i_j$ & Device $j$ in the cluster $C_i$ \\ \hline
        $N_i$ & Total number of devices in Cluster $C_i$ \\ \hline
        $y_n$ & Total received signal at the BS in the $n$-th time slot \\ \hline
         $\mathcal{I}\left(i,j,n\right)$ &  Indicator function that is 1 when $D^i_j$ transmits in the $n$-th time slot, and 0 otherwise \\ \hline
        $\textbf{H}_j^i$ & Complex channel gain vector between $D^i_j$ and BS \\ \hline
        $x_j^i$ & Transmitted signal by device $D^i_j$ \\ \hline
        $W$ & Number of devices that actually transmit in a frame \\ \hline
        $L$ & Total number of time slots in a frame \\ \hline
        $E$ & Device energy consumption \\ \hline
        $a_i$ &Access probability for cluster $C_i$ \\ \hline
        $b_i$ & Seed for cluster $C_i$ \\ \hline
        $\gamma_i$ & Throughput for cluster $C_i$ \\ \hline
        $\gamma_s$ & Total system throughput \\ \hline
        $J(...)$ & Jain's fairness index \\ \hline
        $s$ & System state \\ \hline
        $s_i$ & Total number of successful transmissions in $C_i$ \\ \hline
        $s_{\text{next}}$ & System state in the next frame \\ \hline
        $r$ & Immediate reward received after transition \\ \hline
        %$\pi_i(a_i|s,\vec{\theta}_i)$ & Access probability policy for cluster $C_i$ \\ \hline
        $\pi_i(a_i|s,\vec{\theta}_i)$ & Probability of access probability $a_i$ for state~$s$ \\ \hline
        $\vec{\theta}_i$ & Parameters vector of the access policy for cluster $C_i$ \\ \hline
        %$\tau_i(b_i|s,\vec{\phi}_i)$ & Seed policy for cluster $C_i$ \\ \hline
        $\tau_i(b_i|s,\vec{\phi}_i)$ & Probability of selecting seed $b_i$ in state $s$ for $C_i$ \\ \hline
        $\vec{\phi}_i$ & Parameters vector of the seed policy for cluster $C_i$ \\ \hline
        $\phi^{(j)}_i$ & $j$-th element of the parameter vector $\vec{\phi}_i$ \\ \hline
        $\alpha_{\theta}$ & Learning rate for updating $\vec{\theta}_i$ \\ \hline
        $\alpha_{\phi}$ & Learning rate for updating $\vec{\phi}_i$ \\ \hline
        $\delta$ & Temporal-difference error or learning signal \\ \hline
       % $\nabla\log\pi_i(a_i|s,\vec{\theta}_i)$ & Gradient of log-likelihood of access policy \\ \hline
       % $\nabla\log\tau_i(b_i|s,\vec{\phi}_i)$ & Gradient of log-likelihood of seed policy \\ \hline
        $V(s, \vec{\omega})$ & State-value function at state $s$ with parameters $\vec{\omega}$ \\ \hline
        $\vec{\omega}$ & Parameter vector for value function, dimension $n$ (number of states) \\ \hline
        $\omega^{(s)}$ & $s$-th element of vector $\vec{\omega}$; value estimate for state $s$ \\ \hline
        $\alpha_{\omega}$ & Learning rate for updating $\vec{\omega}$ \\
        \hline
       % $\vec{\phi}_i$ & Parameter vector of dimension $q$ for seed policy $\tau_i$ \\ \hline
        $q$ & Number of candidate seeds for each cluster \\ \hline
      %  $\{b^{(1)}, \dots, b^{(q)}\}$ & Set of all candidate seeds \\ \hline
        $h_i(s, b_i, \vec{\phi}_i)$ & Score function used in the softmax, often set as $\phi_i^{(s)}$ \\ \hline
        %$\alpha_{\phi}$ & Learning rate for seed policy parameter $\vec{\phi}_i$ update \\ \hline
        $e^{h_i(\cdot)}$ & Exponential of the score for softmax normalization \\ 
        \hline     
    \end{tabular} \vspace{-3mm}
    \label{acronym_table}
\end{table}

%\vspace{-3mm}

\justifying
%\raggedright
%\raggedleft
\begin{IEEEbiography} 
[{\includegraphics[width=1in,height=1.25in,clip,keepaspectratio]{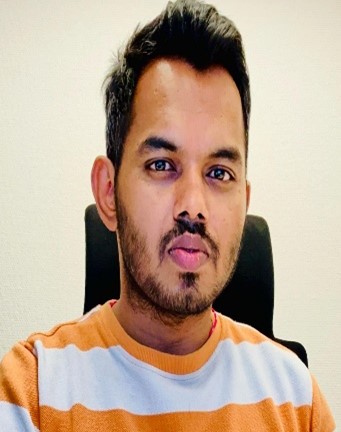}}]{Abhishek Kumar}%(Student Member, IEEE) 
obtained his B.E. degree in Electronics and Communication Engineering from Rajiv Gandhi Proudyogiki Vishwavidyalaya (RGPV), Bhopal, India, in 2017. He further pursued his M.Tech. degree in Electronics and Communication Engineering from the Indian Institute of Technology (IIT) Bhubaneswar, India, in 2021. He is currently working towards a Ph.D. degree at the Department of Information and Communication Technology, University of Agder (UiA), Norway. His research focuses on MAC processes, routing protocols, and resource management in beyond 5G mobile systems, wireless networks, and applied machine learning in IoT networks. 
\end{IEEEbiography}

\vfill

\begin{IEEEbiography}[{\includegraphics[width=1in,height=1.25in,clip,keepaspectratio]{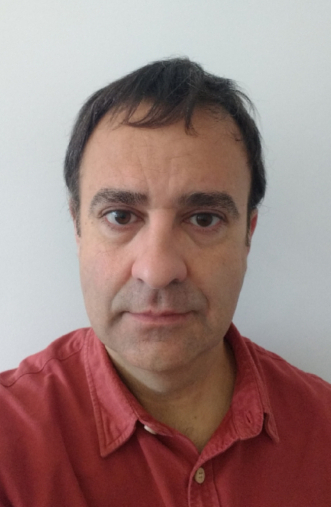}}]
{José-Ramón Vidal} received the Telecommunication Engineering Ph.D. degree from the Universitat Politècnica de València (UPV), València, Spain. He is currently an Associate Professor at the Department of Communicactions, Higher Technical School of Telecommunication Engineering, UPV. His current research interest includes the application of game theory and machine learning to resource allocation in cognitive radio networks and to economic modeling of telecommunication service provision. In these areas, he has authored or coauthored multiple papers in refereed journals and conference proceedings.
\end{IEEEbiography}

\vfill
  
\begin{IEEEbiography}[{\includegraphics[width=1in,height=1.25in,clip,keepaspectratio]{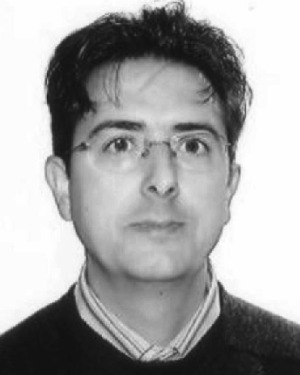}}]{Jorge Martinez-Bauset} received the Ph.D. degree from the Universitat Polit\`ecnica de Val\`encia (UPV), Valencia, Spain, in 1997. He is currently a Professor with the UPV. From 1987 to 1991, he was with QPSX Communications, Perth, Australia, working with the team that designed the first IEEE 802.6 MAN. Since 1991, he has been with the Department of Communications, UPV. His research interests include performance evaluation and traffic control for multiservice networks. He was the recipient of the 1997 Alcatel Spain Best Ph.D. Thesis Award in access networks. 
\end{IEEEbiography}

\vfill

\begin{IEEEbiography}[{\includegraphics[width=1in,height=1.25in,clip,keepaspectratio]{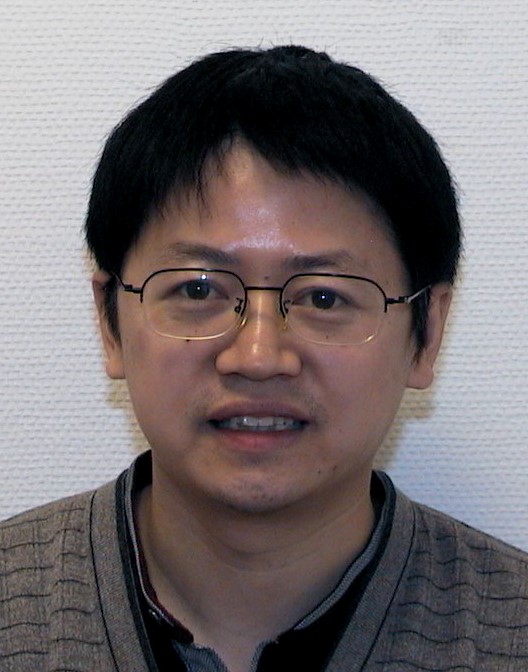}}]{Frank Y. Li} received the Ph.D. degree from the Department of Information Security and Communication Technology, Norwegian University of Science and Technology (NTNU), Trondheim, Norway, in 2003. He was a Senior Researcher at the Department of Technology Systems, University of Oslo, before joining the Department of Information and Communication Technology, University of Agder (UiA), in 2007, as an Associate Professor and then a Full Professor. From Aug. 2017 to Jul. 2018, he was a Visiting Professor with the Department of Electrical and Computer Engineering, Rice University, Houston, TX, USA. During the past years, he has been an active participant in multiple Norwegian and EU research projects. His research interests include MAC mechanisms and routing protocols in beyond 5G mobile systems and wireless networks, Internet of things, mesh and ad hoc networks, wireless sensor networks, D2D communications, cooperative communications, cognitive radio networks, green wireless communications, dependability and reliability in wireless networks, QoS, resource management, and traffic engineering in wired and wireless IP-based networks, and the analysis, simulation, performance evaluation of communication protocols and networks, and applied machine learning in IoT and beyond 5G networks. %He was listed as a Lead Scientist by the European Commission DG RTD Unit A.03- Evaluation and Monitoring of Programmes in Nov. 2007. 
\end{IEEEbiography}

\end{document}